\documentclass{aa}  

\usepackage{graphicx}
\usepackage{txfonts}
\usepackage{soul}
\begin{document}

   \title{MHD simulations of the in-situ generation of kink and sausage waves in the solar corona by collision of dense plasma clumps}

   \subtitle{ }

   \author{P. Pagano\inst{\ref{inst1}} \and
          H.J. Van Damme\inst{\ref{inst1}} \and
          P. Antolin\inst{\ref{inst1}} \and
          I. De Moortel\inst{\ref{inst1},\ref{inst2}}
          }

\authorrunning{Pagano et al.}
\titlerunning{Waves from collision of dense clumps}

\institute{School of Mathematics and Statistics, University of St Andrews, North Haugh, St Andrews, Fife, Scotland KY16 9SS, UK \label{inst1}
            \and
            Rosseland Centre for Solar Physics, University of Oslo, PO Box 1029  Blindern, NO-0315 Oslo, Norway \label{inst2}
 \\
      \email{pp25@st-andrews.ac.uk}
      }

   \date{ }

 
  \abstract
   {MHD waves are ubiquitous in the solar corona where the highly structured magnetic fields provide efficient wave guides for their propagation. While MHD waves have been observed originating from lower layers of the solar atmosphere, recent studies have shown that some can be generated in-situ by the collision of dense counter propagating flows.}
   {In this theoretical study, we analyse the mechanism that triggers the 
   propagation of kink and sausage modes in the solar corona, following the collision 
   of counter-propagating flows and how the properties of the flows
   affect the properties of the generated waves.}
   {In order to study in detail such a mechanism we run a series of ideal 2D and 3D MHD simulations
   where we vary the properties of the counter-propagating flows and by means of a simple technique to estimate the amplitudes of the kink and sausage modes, we investigate
   their role on the generation and propagation of the MHD waves.}
   {We find that the amplitude of the waves is largely dependent on the kinetic energy of the flows and that the onset of kink or sausage modes depends on the asymmetries between the colliding blobs.
   Moreover, the initial wavelength of the MHD waves is associated with the magnetic configuration resulting from the collision of the flows.
   We also find that genuine 3D systems respond with smaller wave amplitudes.}
   {In this study, we present a parameter space description of the mechanism that leads to the 
   generation of MHD waves from the collision of flows in the corona.
   Future observations of these waves can be used to understand the properties
   of the plasma and magnetic field of the solar corona.}


   \maketitle
%

\section{Introduction}
Magnetohydrodynamic waves are an intrinsic feature of the solar corona
and for this reason, have been studied in detail since
the derivation of their existence \citep[for a review see e.g.][]{DeMoortelNakariakov2012,Arregui2015}.
The solar corona is a highly dynamic and structured environment 
where magnetic fields are often concentrated in coronal loops and active regions.
The formation process of coronal loops is still under investigation, but
these magnetised structures are efficient wave guides for MHD waves
\citep{Reale2010}.

This scenario has been observed a number of times and
a photospheric or chromospheric origin is suggested
for most of the MHD waves propagating
along these wave guides,
as these modes are powered by the upward Poynting flux observed at the lower layers of the solar atmosphere
\citep[e.g.][]{2010ApJ...716L..19M,2012ApJ...744...98C,2015ApJ...812L..15K}.
At the same time, \citet{2019NatAs.tmp..196M} have shown that the power spectrum
of the transverse oscillations in the corona is relatively stable and uniform 
throughout different regions and the solar cycle, suggesting that these oscillations must be fundamentally connected with the pulsation of 
the lower layers of the Sun's atmosphere.
However, various observational and theoretical studies  
have recently suggested that not all the transverse waves observed in the solar corona
are of chromospheric origin, but that some are generated in-situ.
For instance, a series of studies 
\citep{2018A&A...613L...3K, 2017A&A...606A.120K, 2017A&A...598A..57V, 2017A&A...601L...2V}
found that coronal rain can be an efficient source for these waves in coronal loops.
When generated at photospheric level, Alfv\'enic waves suffer from strong reflection when travelling through the lower layers of the solar atmosphere. Therefore, a mechanism that allows in-situ wave generation bypasses this problem, and constitutes an additional and potentially important source of energy for the corona.

\citet[][from now on, Paper I]{Antolin2018} have shown how
MHD waves can be generated from the interaction 
of counter propagating flows or clumps of plasma along coronal loops.
It is not clear yet how frequent this mechanism can occur, but
this scenario seems particularly plausible in loops with coronal rain
\citep[e.g.][]{2012ApJ...745..152A,2014ApJ...789L..42K}, in prominences \citep[e.g.][]{2013ApJ...775L..32A} 
or when strong plasma evaporation is present
\citep[e.g.][]{2004ApJ...613..580B,2013ApJ...762..133B,2018ApJ...857..137G}.

In paper I we have focused on the observation and modelling
of a specific event observed on 2014 April 3 during a coordinated observing campaign between the 
Interface Region Imaging Spectrograph \citep[\textit{IRIS, }][]{2014SoPh..289.2733D},
and \textit{Hinode} \citep{2007SoPh..243....3K},
combined with support from the Solar Dynamics Observatory \citep[\textit{SDO},][]{2012SoPh..275....3P},
pointing at a prominence and coronal rain complex
on the West limb of the Sun.
The observations clearly show two rain clumps moving along
the same coronal structure in opposite directions.
When the two clumps intersect in the plane-of-the-sky (POS),
a localised brightening is observed.
At this point, the two clumps merge into one
that continues moving downwards along the same structure,
and a transverse oscillation of the rain strand is observed in the POS.
The analysis of the observations combined with the modelling have led us to interpret
these observations as coupled kink and sausage modes triggered
by the collision of the two clumps.

The model devised for this purpose consisted of an ideal 2D MHD simulation
of two asymmetric clumps travelling towards each other in a uniform medium.
The parameters of the MHD simulations were chosen to match the observations
and the magnetic field strength was picked to retrospectively match the observed kink amplitude.
While the model was suitable to explain one specific observation,
it has triggered new questions about this mechanism
as a way to generate transverse MHD waves in the solar corona.
In particular, the connection between the 
properties of the clumps and properties of the induced transverse modes is still to be clarified, as well as the evolution of the forces during the collision.
A key open issue was in the degree of asymmetry required to trigger kink modes
and whether sausage modes are also a common result.
From a more general perspective, the key question that needs to be answered
in this and future studies is whether the excitation of MHD waves
in the solar corona through flow collision
is a common event or a rather rare phenomenon.
This has important implications for the
energy budget, and the fraction originating from lower layers of the solar atmosphere.
Additionally, this is a potentially very interesting tool for seismology studies.

The main focus of this work is to explain how the interaction of counter propagating 
clumps in coronal loops can lead to the generation of transverse MHD waves.
Also, we aim to explore how the properties of the clumps can influence the properties of the excited kink and sausage modes.
In order to do that, we start from the MHD simulations already presented in Paper I
and we significantly expand that investigation 
by varying the key parameters of the clump collisions.
We also extend the model to 3D and investigate the changes relative to the 2D case. 

The paper is structured as follows.
In Sec.\ref{model} we present the 2D MHD simulations used for our modelling,
in Sec.\ref{simcollision} we analyse in detail how the collision
of two counter propagating clumps generates MHD waves, 
in Sec.\ref{parameterspace} we expand the study by exploring the parameter space of the clump properties,
in Sec.\ref{3dsimulation} we analyse how our results change in a 3D setup,
and we present conclusions in Sec.\ref{conclusions}.

\section{Model}
\label{model}

To study the generation of transverse MHD waves in the solar corona
from counter propagating plasma flows, we have devised a simple MHD model.
This model is capable of reproducing some observed features 
as demonstrated in paper I.
Two dense clumps are set to travel in opposite
directions to reproduce the observed
upward and downward travelling features,
and the properties of the background plasma
and magnetic field
are chosen so that observed properties 
of the transverse waves following the interaction are reproduced.

We repeat here only the key properties of our MHD model and we refer the reader to paper I for further details.
We perform our numerical experiment in a Cartesian 2D domain and we solve
the ideal MHD equations using the MPI-AMRVAC software \citep{Porth2014}.
\begin{equation}
\label{mass}
\displaystyle{\frac{\partial\rho}{\partial t}+\vec{\nabla}\cdot(\rho\vec{v})=0},
\end{equation}
\begin{equation}
\label{momentum}
\displaystyle{\frac{\partial\rho\vec{v}}{\partial t}+\vec{\nabla}\cdot(\rho\vec{v}\vec{v})
   +\nabla p-\frac{\vec{j}\times\vec{B}}{c}=0},
\end{equation}
\begin{equation}
\label{induction}
\displaystyle{\frac{\partial\vec{B}}{\partial t}-\vec{\nabla}\times(\vec{v}\times\vec{B})=0}, 
\end{equation}
\begin{equation}
\label{energy}
\displaystyle{\frac{\partial e}{\partial t}+\vec{\nabla}\cdot[(e+p)\vec{v}]=0} 
\end{equation}
where $t$ is time, $\rho$ the density, $\vec{v}$ velocity,
$p$ the gas pressure, $\vec{B}$ the magnetic field,
$j=\frac{c}{4\pi}\nabla\times\vec{B}$ the current density, 
$c$ the speed of light.
The total energy density $e$ is given by
\begin{equation}
\label{enercouple}
\displaystyle{e=\frac{p}{\gamma-1}+\frac{1}{2}\rho\vec{v}^2+\frac{\vec{B}^2}{8\pi}},
\end{equation}
where $\gamma=5/3$ denotes the ratio of specific heats.

The initial conditions are described in a Cartesian reference frame
where the x-direction is aligned with the initial uniform magnetic field $B=6.5$ $G$.
The simulation domain extends from $x=-6$ $Mm$ to $x=6$ $Mm$ and
from $y=-3$ $Mm$ to $y=3$ $Mm$.
The background plasma has a density of $\rho=10^{-15}$ $g/cm^{-3}$,
corresponding to a number density $n=1.2 \times 10^{9}$ $cm^{-3}$ in a fully ionised plasma
(or an electron density $n_e=6 \times 10^{8}$ $cm^{-3}$),
and a uniform temperature of $T_e=1$ $MK$, leading to a plasma of $\beta=0.098$.
Although coronal rain is not a fully ionised plasma \citep{2012ApJ...745..152A},
we do not expect this assumption to have a
significant effect on the study carried out here.
Indeed, \cite{2016ApJ...818..128O} found that even in the extreme case of 50\% partially ionised plasma, the neutrals are still strongly coupled to the ions.
The clumps are in pressure equilibrium with the background,
but have a density $\rho_C$ times larger and thus a temperature of
$T=T_e/\rho_c$ $K$. The clumps are travelling at a speed $V_C$ in opposite directions.
Each clump has a trapezoidal shape oriented so that 
the two opposite faces are parallel to one another
with an angle $\theta$ of inclination 
between these faces and the
direction perpendicular to the direction of travel
(i.e. y-direction).
The initial distance of the clumps is $d=3$ $Mm$.

The numerical experiment runs for $300$ $s$ which is sufficiently long 
for the collision between the clumps to occur and for 
the system to approach a new equilibrium in the aftermath of the collision.
The boundary conditions are treated with a system of ghost cells 
where we set zero gradient for all
MHD variables at both $x$ and $y$ boundaries.

\section{Simulation}
\label{simcollision}

\begin{figure}
\centering
\includegraphics[scale=0.25]{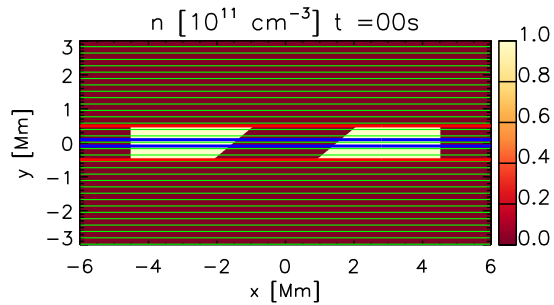}
\caption{Map of the number density $n$ at $t=0$ $s$.
Green lines are magnetic field lines drawn from the left-hand-side boundary.
Blue and red lines are specific external (red) and internal (blue)
pairs of magnetic field lines that are used in Section \ref{ksmodes}.}
\label{initcond}
\end{figure}

As the main focus of this work is to describe the interaction of counter propagating clumps in a coronal loop, in this section we illustrate in detail one of the characteristic simulations presented in Paper I.
This section significantly expands the analysis described in Paper I and it is essential to introduce the subsequent parameter space investigation.
In particular, we will illustrate in detail the mechanism through which MHD waves are generated in our numerical experiment.
Fig.~\ref{initcond} shows the initial condition of this numerical experiment
where two dense clumps are aligned and travelling
with the same speed in opposite directions.
In this representative simulation, two clumps of trapezoidal shapes
are set to travel towards each other at a speed of $v_C=70$ $km/s$.
The two clumps have a density contrast $\rho_C=100$,
their width is $w=1$ $Mm$ , their length at the centre is $L=3$ $Mm$,
and the angle of the facing clumps surfaces is $\theta_C=50^{\circ}$.
This setup was found to produce the best match
with the observed dynamics (paper I).

\subsection{Collision}
\label{collision_beg}

\begin{figure*}
\centering
\includegraphics[scale=0.14]{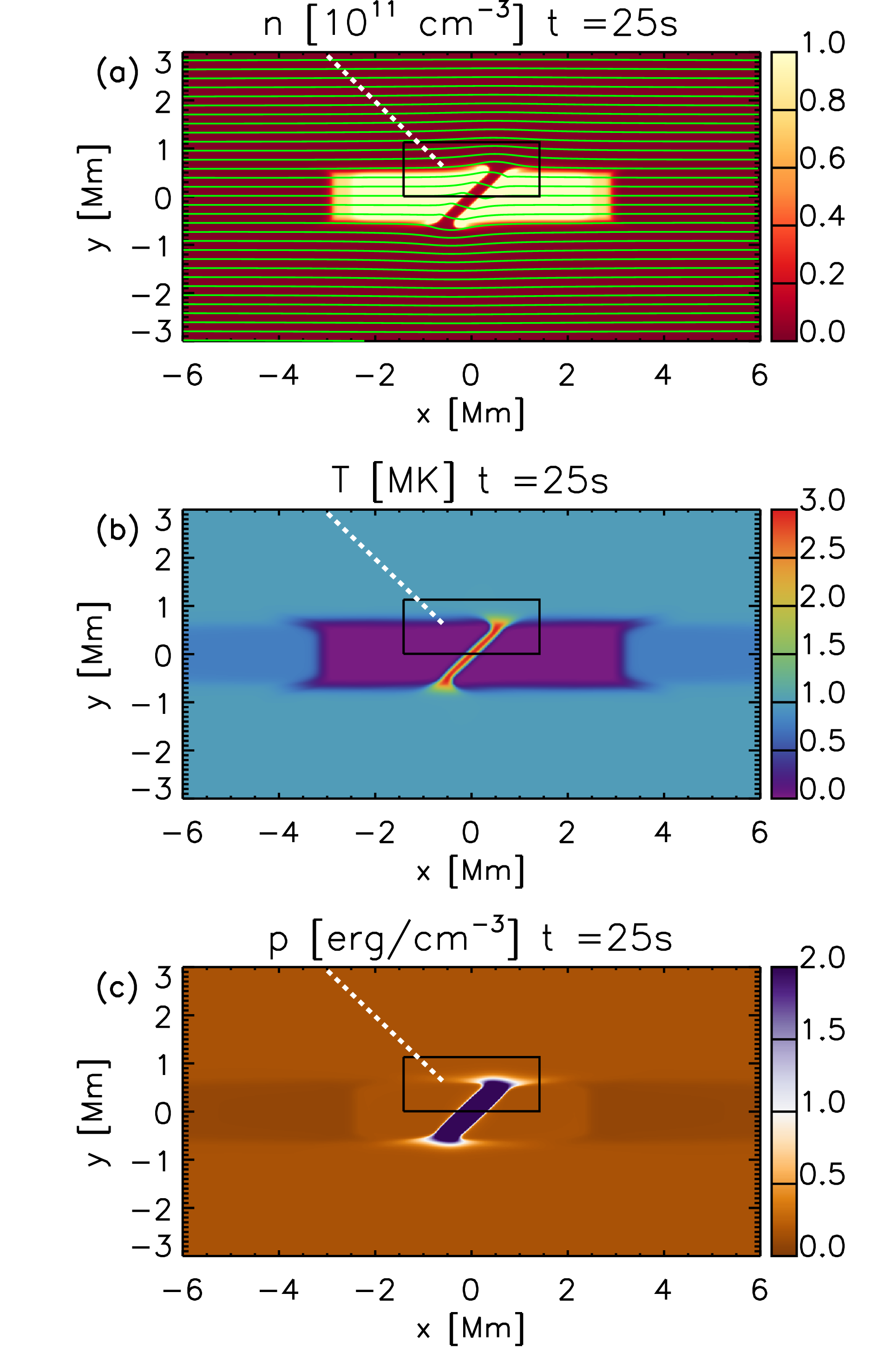}
\includegraphics[scale=0.14]{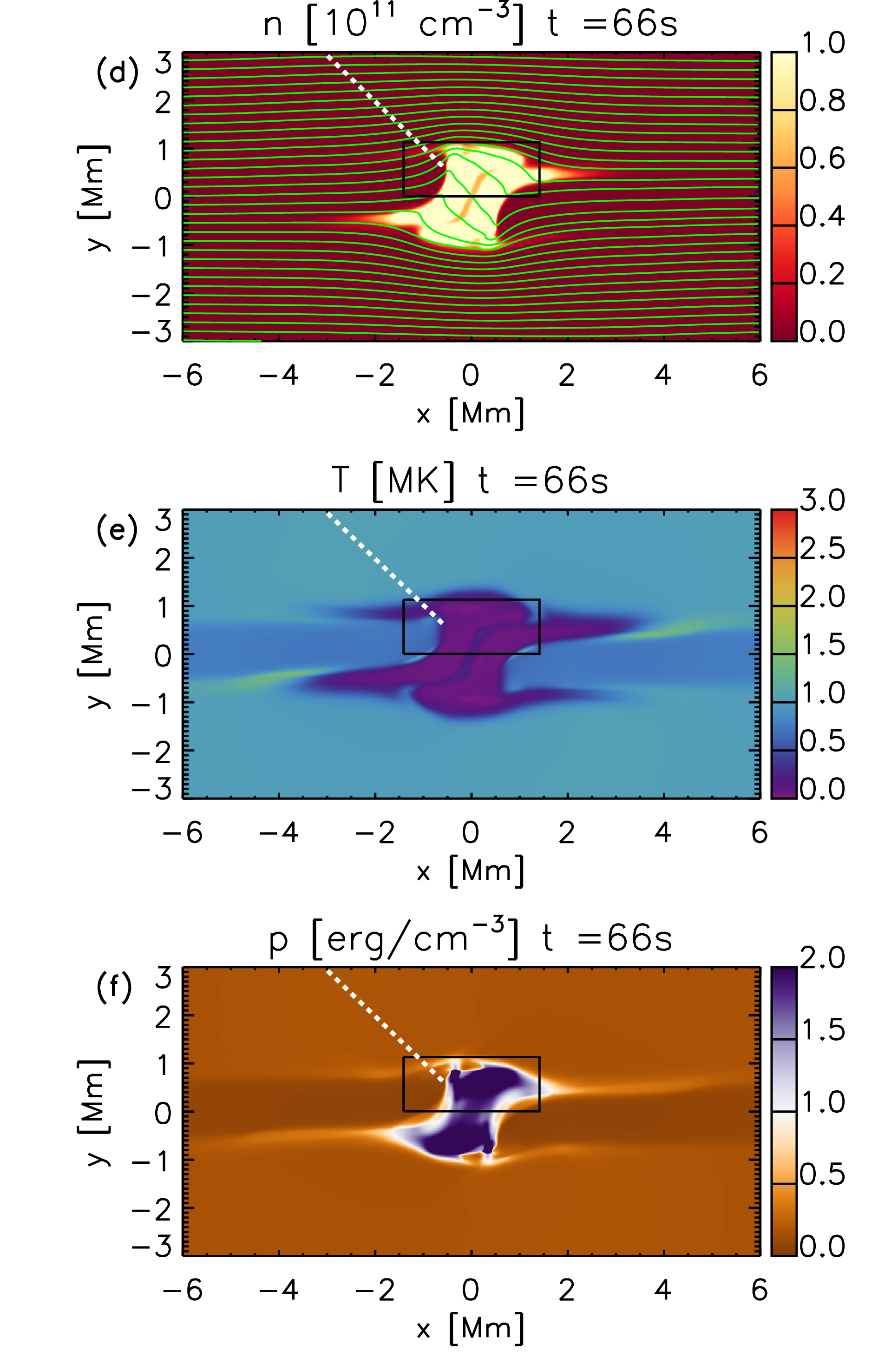}
\includegraphics[scale=0.14]{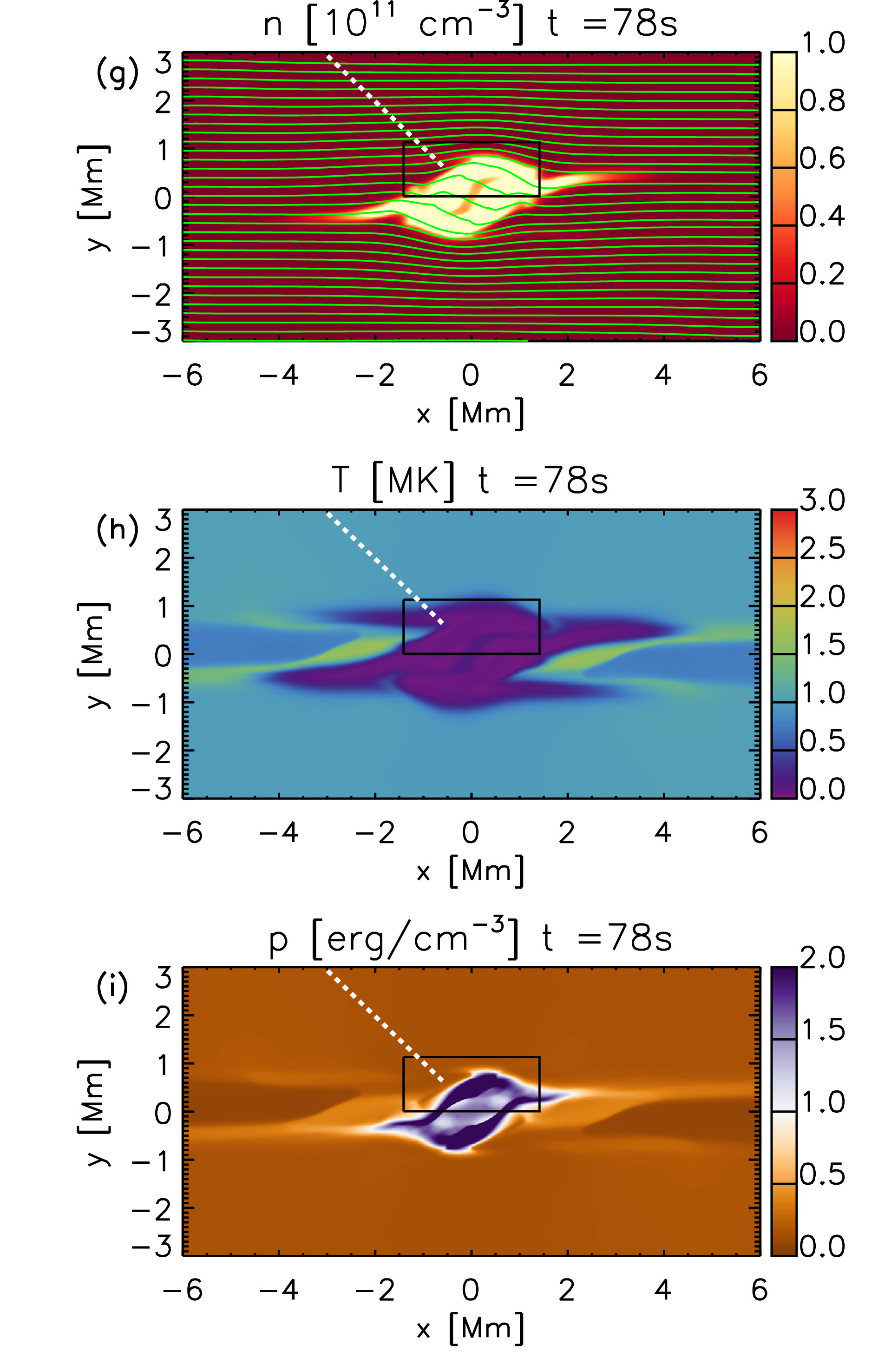}
\caption{Maps of number density, temperature $T$ and gas pressure $p$ at different times 
of the MHD simulation.
In the $\rho$ maps, green lines are magnetic field lines drawn from the left-hand-side boundary.
The rectangular box shows the region over which we average quantities for Fig.\ref{forces}
and the white dashed line is the cut we use to plot the time distance diagram in Fig.\ref{fastmodes}.
Movie is available online.}
\label{snaps}
\end{figure*}

\begin{figure}
\centering

\includegraphics[scale=0.30]{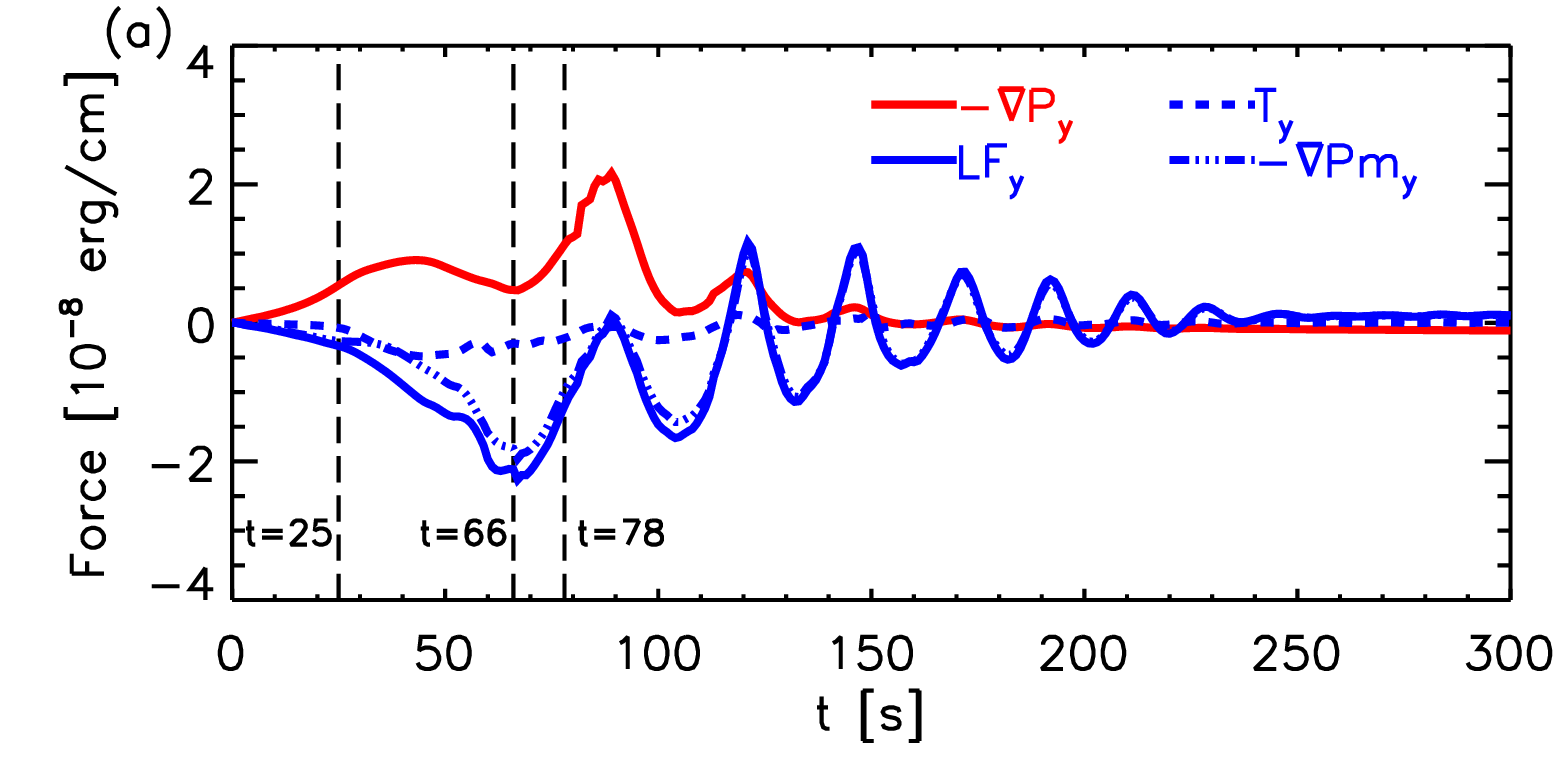}

\includegraphics[scale=0.30]{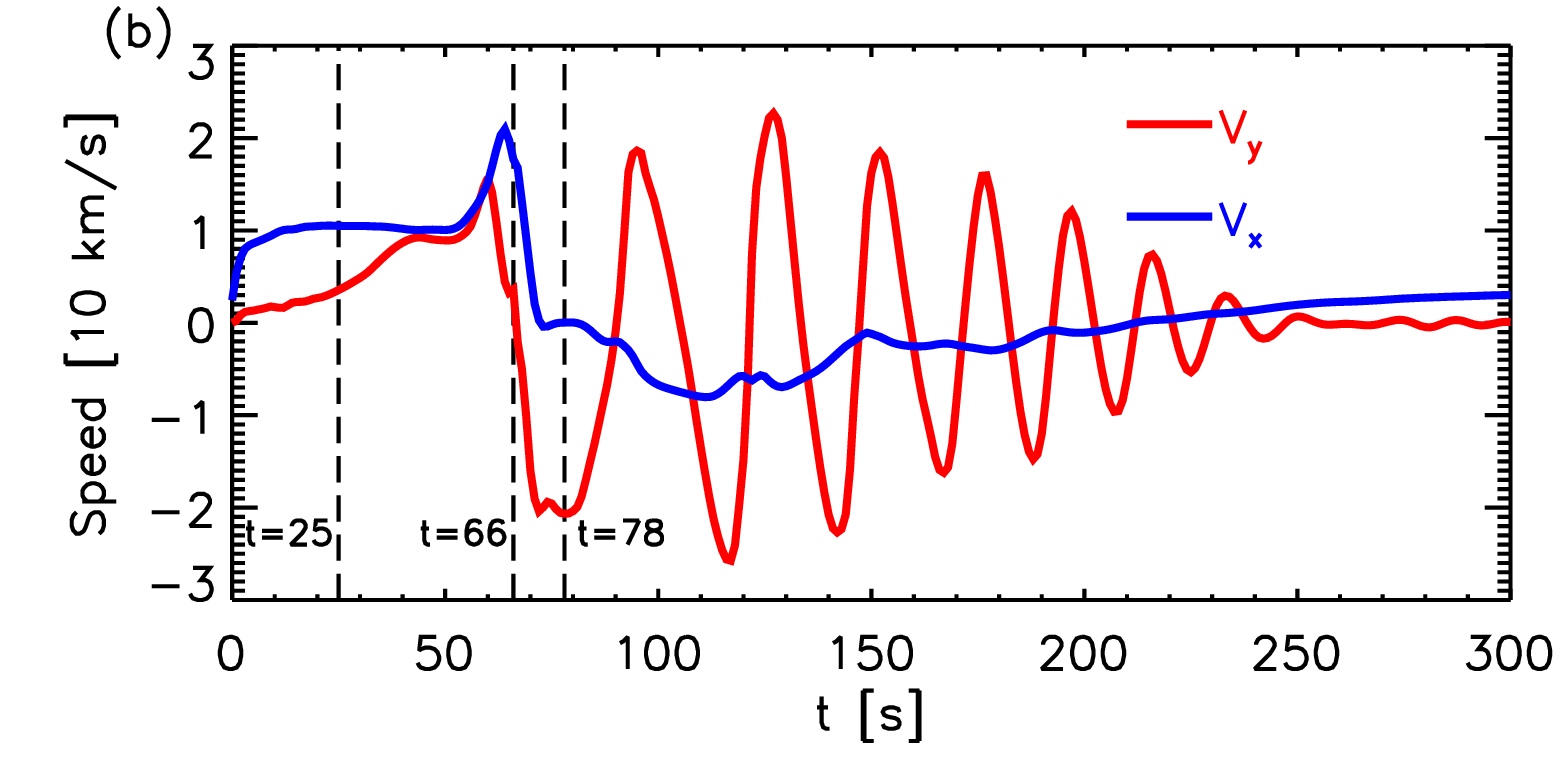}

\caption{(a) Evolution of the $y$-components of the gas pressure gradient ($-\nabla P_{y}$, red solid curve), Lorentz force ($LF_{y}$, blue solid curve),
magnetic tension ($T_{y}$, blue dashed curve) and magnetic pressure gradient ($-\nabla Pm_{y}$, blue dash-dotted curve) averaged over the rectangular region shown in Fig.~\ref{snaps}.
(b) Evolution of the $x$ and $y$ components of the velocity  averaged over the rectangular region shown in Fig.~\ref{snaps}.}
\label{forces}
\end{figure}

The initial speed of the clumps relative to the background medium is
lower than both the Alfv\'en speed ($V_A=600$ $km/s$)
and the sound speed ($C_s=165$ $km/s$), thus no shock initially develops
and the two clumps travel at constant speed in opposite direction.
Doing so, their motion leads to a compression of the plasma ahead of their propagation
and a rarefaction behind.

At $t=7$ $s$ the two compressed regions touch and the collision phase begins.
Figs.~\ref{snaps}a-c show $\rho$, $T$, and $p$ of the
system at $t=25$ $s$, an intermediate stage during the collision.
At this time the clumps have started to change shape at their fronts because
of the interaction.
In particular, the front of the clumps has expanded in the y-direction and the 
magnetic field lines have followed this deformation.
It should be noted that the magnetic field is most deformed where the two clumps surfaces
are touching, thus at different x-coordinates for the upper and lower sides of the clumps
because of their asymmetric shape.
At this stage, the compression between the clumps has caused a significant increase in both pressure and temperature. The latter reaches about $T=3$ $MK$, which is its maximum over the course of the simulation.
This high temperature is maintained only for a couple of seconds before it drops again.
At this time, the gas pressure distribution between the clumps follows
the pattern of the two colliding fronts and strong gradients are generated along the collision region.
Over the same region, the y-velocity of the plasma is outward directed ($v_y\sim\pm10$ $km/s$) with respect to the collision and the magnetic field is displaced with the plasma.

At this initial stage of the collision, the still unbalanced increase of the gas pressure between the clumps drives the dynamics.
Fig.~\ref{forces}a shows the evolution of the average forces as a function of time 
in the black rectangle shown in Fig.\ref{snaps} that
covers the $y$-positive part of the collision region.
In this rectangle positive and negative forces are directed outwards and inwards, respectively, from the centre of the clumps along the direction perpendicular to their motion.
The gas pressure gradient in this region remains always outward directed and 
in this phase ($t<40$ $s$), it is about 30\% stronger
than the Lorentz force.
At this stage the magnetic field deformation is still minimal and the magnetic tension force is a minor component of the total Lorentz force.

After the collision starts, the distortion of the magnetic field just outside of the clumps immediately generates forces directed inwards
that resist any further expansion of the clumps in the y direction and
limits the distortion of the magnetic field.
At the same time, as the clumps have a finite extent, the plasma coming 
into the collision region keeps the initial dynamic ongoing
and so the magnetic field distortion continues.
However, in this system the restoring Lorentz force exerted by the magnetic field is proportional to the departure from the initial equilibrium condition
and it overcomes the gas pressure gradient over time.
This happens at $t=40$ $s$ for the average forces,
as they become inward directed.
However, the ongoing collision and the inertia of the plasma continue
the distortion of the frozen in magnetic field as can be seen in Fig.\ref{forces}b
where the average $v_x$ and $v_y$ components of the velocity are shown.
We find that as the collision progresses, the average value of $v_y$ keeps growing
until $t=60$ $s$ when it reaches a maximum and
only at this point does the restoring forces start decelerating the plasma.
Then, $v_y$ remains positive for 6 seconds more,
and at $t=66$ $s$
the forces finally invert the plasma motion and the Lorentz force starts decreasing,
leading to the relaxation of the magnetic field.

Thus, at $t=66$ $s$ (Figs.\ref{snaps}d-f),
the maximum magnetic field distortion is found and the restoring forces start driving the dynamics, leading to the excitation and propagation of MHD waves along the waveguide.
In this system, the time elapsed between the beginning of the collision and the moment when the waves are released depends on i) the duration of the clumps collision, that is of course proportional to their length and ii) the speed of the clumps as their inertia extends the collision phase.
During the collision phase, the magnetic field starts showing a kink-like distortion due to the oblique angle of the colliding interfaces,
i.e. two opposite peaks in the traced magnetic field lines, asymmetrically placed along $x$.
The two peaks are symmetrically placed along $x$ at first,
but they travel apart over the collision time.
When the collision finishes,
it is this asymmetric magnetic configuration that determines the initial wavelength of the kink waves.

The average $x$-component of the velocity 
in the rectangular region shows a similar evolution over the course of the collision ($t<78$ $s$ in Fig\ref{forces}b). It is initially positive, it decreases after $t=66$ $s$ and it becomes negative at $t=78$ $s$.
This evolution lags behind the one of $v_y$,
demonstrating that the plasma is pushed back (negative $v_y$) because 
the restoring magnetic forces push it inward.
The corresponding rectangle in the lower side of the domain gives instead an
opposite average $v_y$ and $v_x$ evolution.

As soon as the collision terminates, the restoring magnetic forces act to bring the system back to equilibrium.
However, this mechanism triggers oscillations
and the density distribution changes 
because of the waves propagating in the system.
Figs.\ref{snaps}g-i show the system when the highest $y$-velocities are found after the collision,
that is when restoring forces have induced a plasma motion opposite to the one initially induced by the clumps' collision.
At this point, the plasma velocity in the $y$-direction 
is directed inward (Fig.\ref{forces}b) and a new regime sets in.

After $t=78$ $s$, when the inertial force from the collision ends and the magnetic field
is no longer driven by plasma motion, the system fundamentally can be regarded as a wave guide
which contains an initial perturbation.
After this time, the system starts oscillating around a configuration not far from equilibrium . This is visible in Fig.~\ref{forces}, where 
the average Lorentz force (the restoring force) and the average $y$-component of the velocity oscillate out of phase by $\pi/2$.

\begin{figure}
\centering

\includegraphics[scale=0.30]{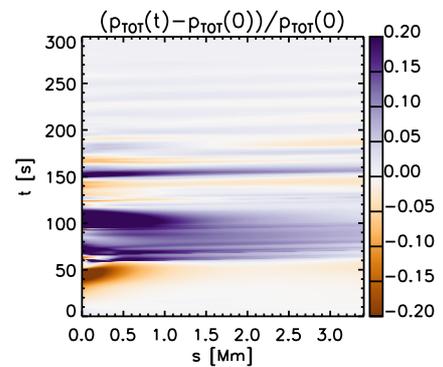}

\caption{Time distance diagram of the total pressure along the cut shown in Fig.\ref{snaps}.}
\label{fastmodes}
\end{figure}

Throughout this evolution, there are also clear signatures of fast MHD modes propagating away from the collision region when the collision ends and the total force exerted on the clumps becomes inward directed.
This is shown in Fig.~\ref{fastmodes}, where a time distance map of the total pressure variations
with respect to the initial one are shown along the cut represented by the dashed line in Fig.~\ref{snaps}.
Distance is represented from the location closer to the centre of the domain.
We find that after the collision time ($t=60$ $s$) signals 
of total pressure increase (compression) travel away from the collision at the local Alfv\'en speed in what constitutes a train of fast MHD waves.
Although this MHD simulation does not have the required output cadence to study in detail
the propagation of these signals, it is worth mentioning their existence
as they are a potential observational signature for the detection of clump collisions in the solar corona.

\subsection{Analysis of kink and sausage modes}
\label{ksmodes}

\begin{figure}
\centering
\includegraphics[scale=0.20]{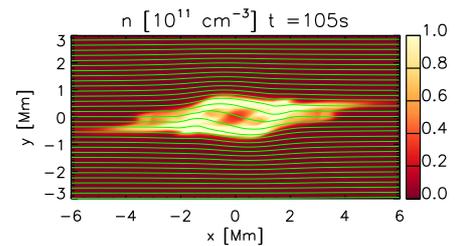}
\caption{Map of the number density $n$ at $t=105$ $s$.
Green lines are magnetic field lines drawn from the left-hand-side boundary.}
\label{snap105}
\end{figure}

In this numerical experiment, the asymmetric fronts
of the colliding clumps cause a distortion of 
the magnetic field that resembles a kink wave propagating
along a waveguide.
Similarly, we also find symmetric expansion and compression of the waveguide that resembles sausage modes.
Fig.~\ref{snap105} shows a state of the MHD simulation where it is possible
to visually identify the presence of both the kink mode distortion and sausage mode expansion 
of the wave guide.
Based on this, we develop a simple technique in order to identify the 
kind of wave modes induced and whether kink or sausage modes are predominantly present.

\begin{figure}
\centering
\includegraphics[scale=0.40]{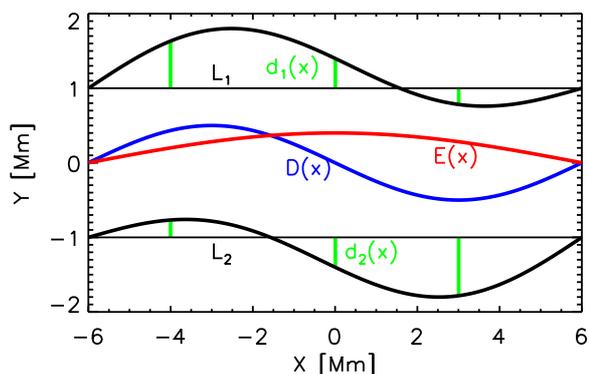}
\caption{Sketch to show the distortion of a pair of magnetic field lines
$L_1(x)$ and $L_2(x)$ can be interpreted in terms of the displacements $d_1(x)$ and $d_2(x)$
to derive the functions $D(x)$ and $E(x)$.}
\label{modes}
\end{figure}

Fig.\ref{modes} illustrates how this technique works.
Let us consider any pair of initially straight magnetic field lines $L_1(x)$ and $L_2(x)$
that are placed symmetrically about the central axis of a magnetic wave guide.
At a given time, after the wave guide is perturbed, we measure the displacement
from their original position
$d_1(x)$ and $d_2(x)$
for both magnetic field lines.
We then combine these displacements to derive the functions:
\begin{equation}
    D(x)=\frac{1}{2}\left(d_1(x)+d_2(x)\right)
\label{Dd1d2}
\end{equation}
\begin{equation}
    E(x)=\frac{1}{2}\left(d_1(x)-d_2(x)\right)
\label{Ed1d2}
\end{equation}
Fig.~\ref{modes} shows the profile of $D(x)$ and $E(x)$ for a given
pair of $d_1(x)$ and $d_2(x)$.
The function $D(x)$ follows the bulk displacement of the wave guide,
as it is proportional to the coherent displacement of $L_1$ and $L_2$ and it is zero
when they move symmetrically in opposite directions.
In contrast, $E(x)$ is non zero when the wave guide expands or contracts.
Following this approach, we consider $D(x)$ indicative of kink-like oscillations in the wave guide
and $E(x)$ of sausage-like ones.
Applying this procedure to a given pair
of magnetic field lines 
for all snapshots of our simulation 
and taking the maximum of $D(x)$ and $E(x)$ for each of them,
we obtain how the kink and sausage modes evolve during the simulation.
It is important to note that the waveguide in our MHD simulation
is highly variable in time and structured in space
and therefore, the properties of the trapped kink and sausage modes 
are equally complex.
Thus, we do not attempt to measure the total energy of these modes.
Still, ours is an estimate of the amplitude of the waves and it can be helpful for comparisons with observed wave amplitudes that are equally complex measurements.
Moreover, we are able to analyse whether some modes are likely to strengthen or weaken in time and whether their relative importance increases or decreases in connection with other parameters.

\begin{figure*}
\centering
\includegraphics[scale=0.40]{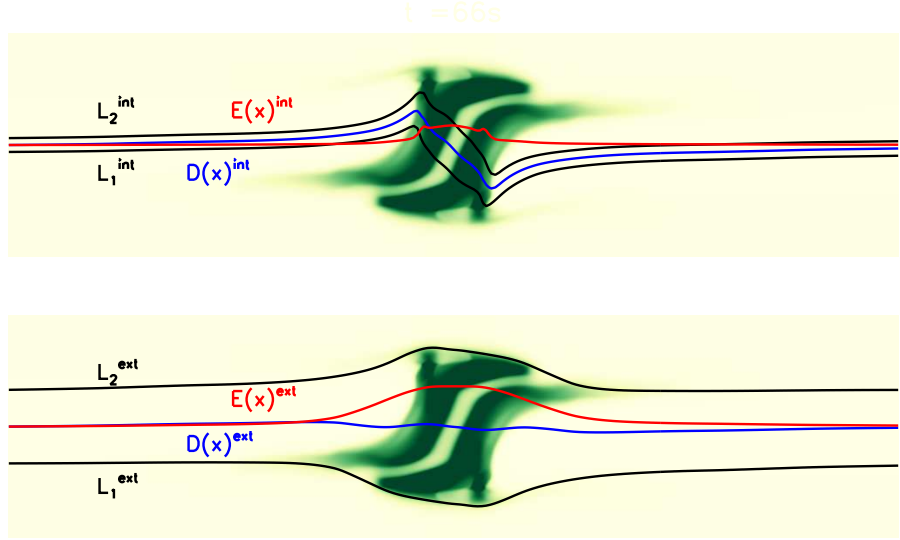}
\caption{Configuration of two pairs of magnetic field lines,
$L_{1}^{int}$ and $L_{2}^{int}$ in the upper panel,
and $L_{1}^{ext}$ and $L_{2}^{ext}$ in the lower panel
at $t=66$ $s$}
\label{fidllines66}
\end{figure*}

In our MHD simulation, we apply this procedure taking into consideration two different pairs of magnetic field lines. 
The first pair ($L_{1}^{ext}$ and $L_{2}^{ext}$) are magnetic field lines initially located at the y-edges of the clumps (red lines in Fig.\ref{initcond}).
These magnetic field lines are useful to study the reaction of the background field to the clump collision.
The second pair ($L_{1}^{int}$ and $L_{2}^{int}$)
are magnetic field lines close to the $y=0$ axis (blue lines in Fig.\ref{initcond}) and are therefore internal to the clumps.
Fig.~\ref{fidllines66} illustrates this analysis at $t=66$ $s$,
when we find the highest value for $D(x)$ for $L_{1}^{int}$ and $L_{2}^{int}$,
and it also corresponds to
when the internal magnetic field lines show the largest distortion.

\begin{figure}
\centering
\includegraphics[scale=0.40]{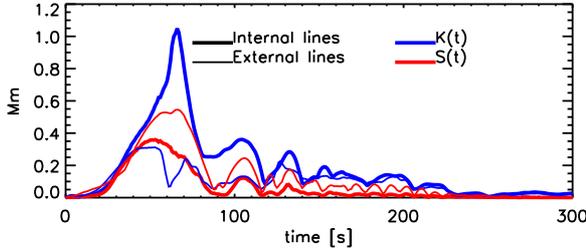}
\caption{Evolution of the functions $K_i(t)$, $K_e(t)$, $S_i(t)$, and $S_e(t)$
that we use to estimate the amplitudes of the kink and sausage modes in the MHD simulation.}
\label{modestime}
\end{figure}

We also define $K_i(t)$ and $K_e(t)$ as the difference between the maximum and minimum values of $D(x)$ at each time $t$ for the internal and external field lines, respectively, and $S_i(t)$ and $S_e(t)$, the corresponding difference between the maximum and minimum values of $E(x)$.
Fig.~\ref{modestime} shows the evolution of these functions in our MHD simulation.
If we focus on the internal magnetic field, we find that $K_i(t)$
rapidly increases in the first phase of the evolution when the magnetic field
undergoes a significant distortion due to the collision of the clumps.
It then quickly decreases after the end of the collision and it undergoes a couple of 
oscillations.
$S_i(t)$ shows a very similar pattern as $K_i(t)$ after the collision,
except that it remains always smaller than $K_i(t)$.
However, during the collision the ramp up and decay phase are more gradual.
The relative amplitude between $K_i(t)$ and $S_i(t)$ is representative
of the driving dynamics within the collision
where the asymmetry of the colliding clumps leads to a magnetic field 
distortion, so that the wave guide is displaced and kinked, but it 
undergoes only a marginal expansion.
In contrast, for the external field lines, $S_e(t)$ is usually larger than $K_e(t)$,
and $S_e(t)$ follows a very similar pattern as $K_i(t)$ initially.
This happens because the external magnetic field lines react to the ongoing collision at this stage
and they expand close to the middle of the domain where the pressure gradient becomes stronger.
$S_e(t)$ is also the quantity that shows the longest lasting oscillatory behaviour 
which shows that after the collision terminates and the kink waves propagates out of the domain,
the wave guide mostly shows sausage oscillations that set a standing wave.
The last visible oscillations of $S_e(t)$ (around $t=200$ $s$)
show a period of about 20 seconds, which is consistent with
the transverse Alfv\'en time in the wave guide. Similar conclusions were drawn by \citet{2004MNRAS.349..705N} 
and \citet{2018ApJ...855...53L}.

In this scenario, we select the functions that show larger amplitudes to 
represent our kink and sausage modes, as they would be predominantly observed.
So, we use $K(t)=K_i(t)$ to represent the kink oscillation induced
by the collision of the clumps that is generated within the collision region,
and $S(t)=S_e(t)$ to represent the induced sausage mode oscillations 
that are mostly generated by the reaction of the background field against the expansion of the wave guide.

When we focus on the oscillations that follow after the end of the collision,
we find from Fig.~\ref{modestime} that two peaks of the kink oscillations
are about $32$ $s$ apart (between $t=100$ $s$ and $t=132$ $s$)
which would be half of the period of an oscillation.
From the spatial profile of $K(t=105)$ and $K(t=132)$ (not shown here),
we can estimate a travel distance of the peaks of about $4.7$ $Mm$.
These estimations give a velocity of about $73$ $km/s$ that corresponds to the Alfv\'en speed
inside the wave guide after the collision,
where we find $|B|\sim6$ $G$ and $\rho=5\times10^{-14}$ $g/cm^3$.

\section{Parameter Space investigations}
\label{parameterspace}


In this section, we aim to understand how the collision mechanism is affected by some basic properties of the clumps. For this, we conduct a parameter space investigation taking as parameters the densities, velocities, lengths and widths of the clumps, as well as the angle and offset between the colliding clumps.

\subsection{Wave amplitude}
\label{waveamplitude}

We first investigate how the strength of the collision affects the 
wave amplitude. As mentioned in paper I,
the background plasma $\beta$ is a key parameter that affects
the amplitude of the induced kink in the wave guide as the magnetic field
tension is stronger when the plasma $\beta$ decreases. 
The sausage mode amplitude increases as well when $\beta$
increases, however we find that kink mode amplitudes
are roughly affected twice as much by the $\beta$ variation relative to the sausage mode.
Here, we keep the background properties unchanged and we vary the 
density contrast and the velocity of the two clumps.
We run 25 numerical experiments in a
$5\times5$ grid of the two parameters we investigate.
The reference simulation is the one presented in Sec.~\ref{simcollision}
($\rho_c=100$, $v_C=70$ $km/s$) and we then 
run numerical simulations 
with $v_C=[70/2,70/\sqrt{2},70\times\sqrt{2},70\times2]$ $km/s$
and $\rho_C=[100/2,100/\sqrt{2},100\times\sqrt{2},100\times2]$.

\begin{table}[]
\begin{tabular}{|l|l|l|l|l|l|}
\hline
$\rho_C$  & {50} & {71} & {100} & {141} & {200} \\ \cline{1-1}
$v_C$     &                     &                       &                      &                      &                      \\ \hline
35 km/s   & 0.06                & 0.10                  & 0.18                 & 0.32                 & 0.52                 \\ \hline
55 km/s & 0.17                & 0.31                  & 0.50                 & 0.73                 & 1.03                 \\ \hline
70 km/s   & 0.49                & 0.71                  & 1.05                 & 1.43                 & 1.79                 \\ \hline
99 km/s   & 1.01                & 1.48                  & 1.91                 & /                    & /                    \\ \hline
140 km/s  & 1.98                & 2.49                  & /                    & /                    & /                    \\ \hline
\end{tabular}
\caption{Table that summarises the density contrast (vertically) and velocity (horizontally)
of the clumps in a parameter space investigation.
The number in each cell represents the maximum of the function $K(t)$ in $Mm$ in each simulation. For some simulations the magnetic field distortion was too disruptive to measure $K(t)$.}
\label{tablerhocvc}
\end{table}
Table~\ref{tablerhocvc} reports the maximum kink amplitude
for each simulation in this investigation.
It should be noted that the measurements of the kink amplitude used in
paper I followed a slightly different technique,
resulting in a value of $\sim1.10$ $Mm$ instead of $1.05$ $Mm$
for the central simulation in Table\ref{tablerhocvc}.
We do not consider this difference critical for the results of this study.
In the simulations where no kink amplitude is reported,
the collision is significantly more disruptive 
than what we have analysed so far
and the magnetic field distortion is too large to be analysed in terms of
the excitation of kink or sausage modes.
For instance, in these simulations, the guide magnetic field
becomes so distorted that $B_x$ changes sign during the collision.
The structure of this grid of experiments is such that
two cornering experiments have the same initial momentum
($\propto\rho_C v_C$), and 
similarly, experiments with the same initial kinetic energy
($\propto\rho_C v_C^2$)
are always two cells in $\rho_c$ and one cell in $v_C$ apart.
\begin{figure}
\centering
\includegraphics[scale=0.40]{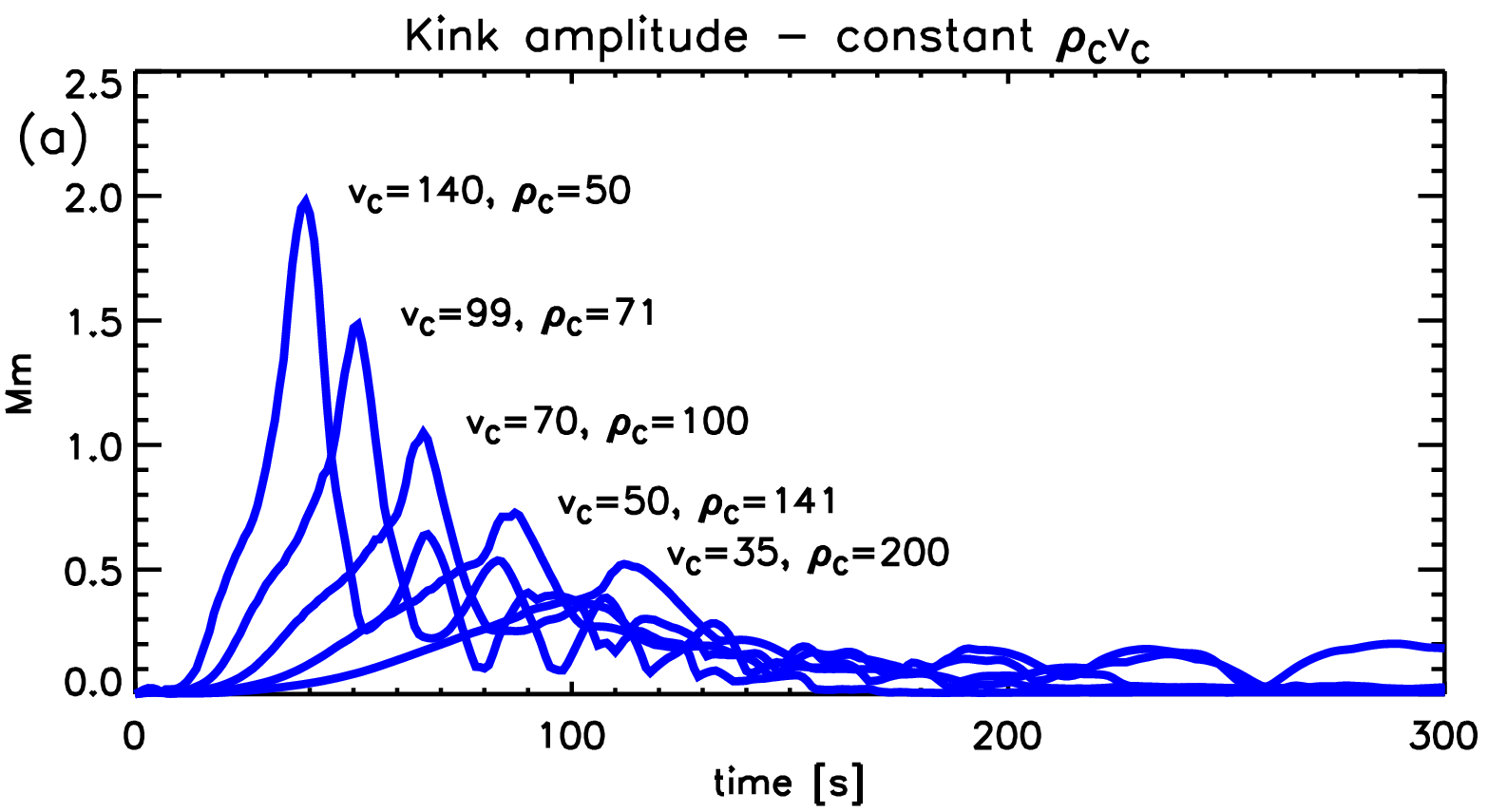}
\includegraphics[scale=0.40]{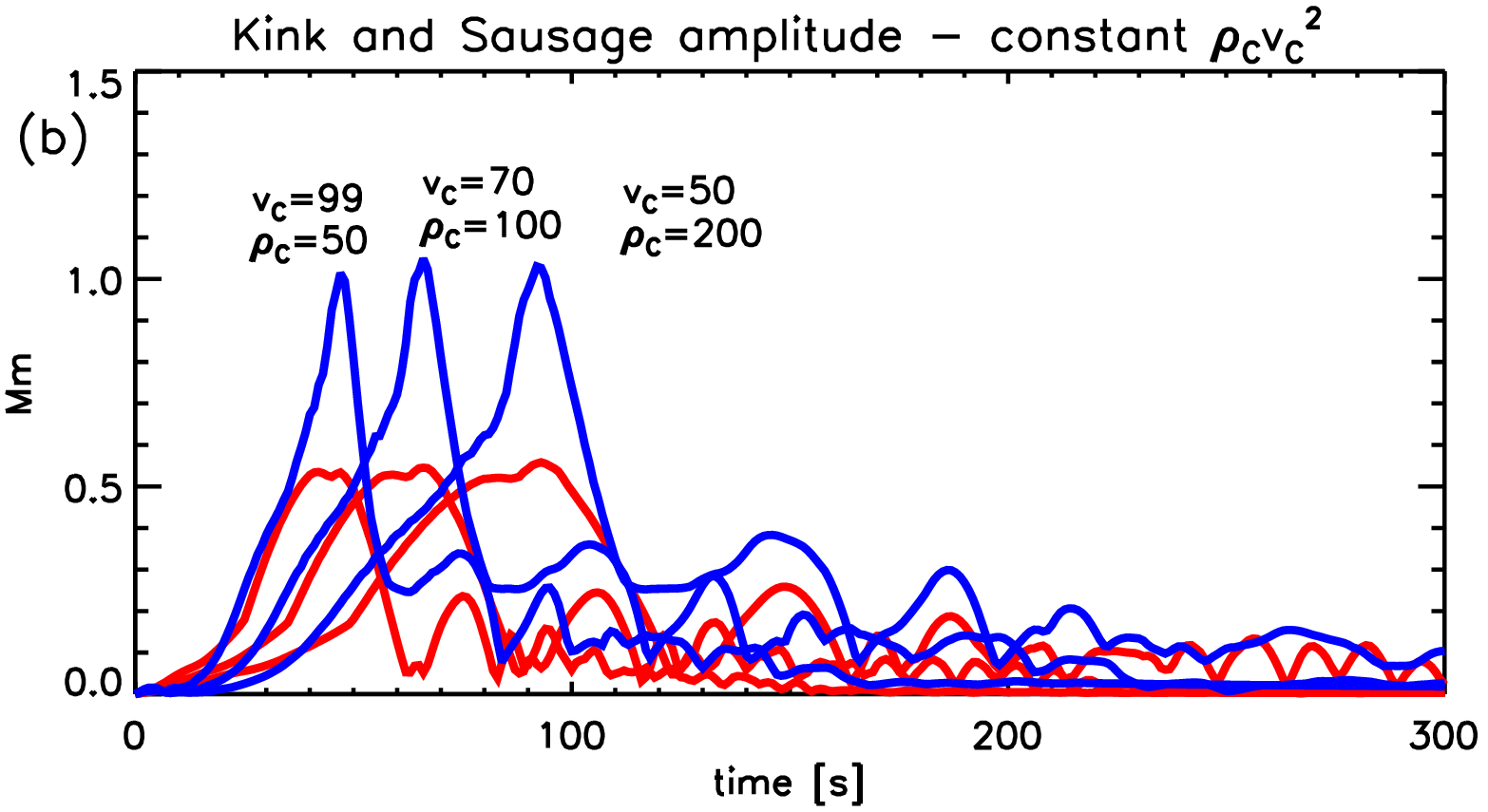}
\caption{(a) Evolution of the function $K(t)$
for simulations with constant initial clumps momentum along the diagonal in Tab.\ref{tablerhocvc}.
(b) Evolution of the functions $K(t)$ (blue lines) and $S(t)$ (red lines)
for simulations with constant initial kinetic energy in Tab.\ref{tablerhocvc}.}
\label{modestime_rhocv}
\end{figure}
Fig.~\ref{modestime_rhocv} shows how the kink amplitude evolves in time
for some numerical experiments with constant momentum (Fig.~\ref{modestime_rhocv}a)
and for some with constant kinetic energy (Fig.~\ref{modestime_rhocv}b).
Fig.~\ref{modestime_rhocv}a clearly shows that the same collision momentum does not lead to
similar kink amplitudes,  and that the velocity plays a dominant role.
The difference in time of maximum kink amplitudes (in both panels in the figure) is just due to the fact that faster clumps lead to earlier collisions.
However, the peak time does not linearly depend on the travelling time of the clumps.
Faster travelling clumps lead to a delay of the times at which maxima are reached.
This happens because these clumps need a longer deceleration time before the magnetic forces
invert the plasma motion.
In contrast, the simulations displayed in Fig.~\ref{modestime_rhocv}b
show very similar maximum kink and sausage amplitudes.
These experiments show that the amplitude of the waves generated by the collisions of counter propagating
clumps essentially depends on their kinetic energies.
This implies that the efficiency of the mechanism
that converts the initial kinetic energy into wave energy is a function of the kinetic energy.
Indeed, as can be seen in Fig.~\ref{modestime_rhocv}b, the energy distribution between the kink and sausage modes
does not seem to be affected by the contrast in density and velocity between the clumps either, as long as the kinetic energy of the collision is maintained. 

\begin{figure}
\centering
\includegraphics[scale=0.30]{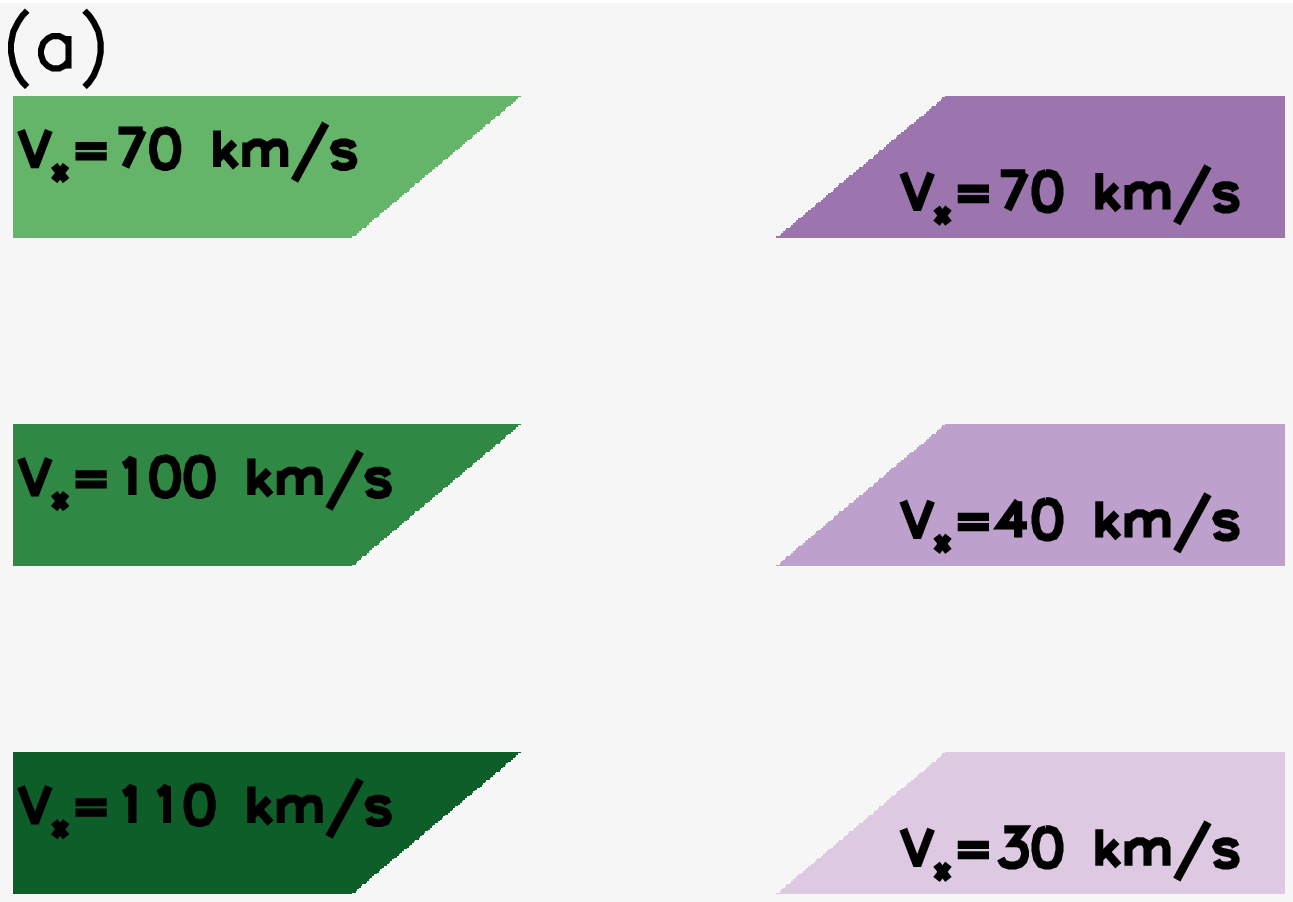}
\includegraphics[scale=0.25]{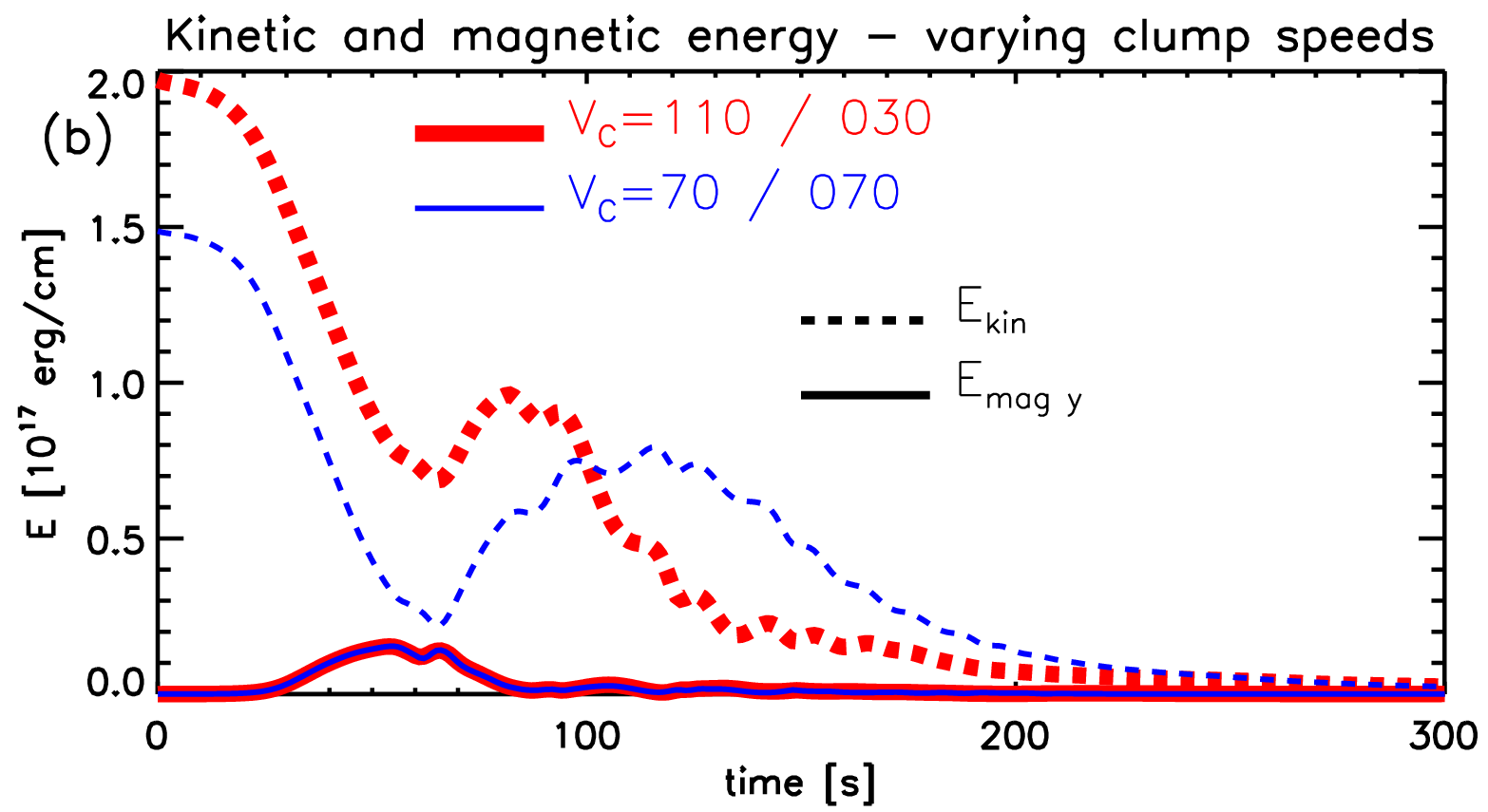}
\caption{(a) Sketch to illustrate some simulations of the parameter space investigation where the speeds between the clumps are varied.
(b) Evolution of the kinetic and magnetic energy for the simulations where the clump speeds are both 
$70$ $km/s$ and where they are $110$ $km/s$ and $30$ $km/s$.}
\label{mapsasymv}
\end{figure}

It should be noted that the key parameter
is the kinetic energy with respect to the centre of mass of the system $E_{K(CM)}$,
\begin{equation}
E_{K(CM)}=\frac{1}{2}\rho\left(\vec{v}-\vec{V_{CM}}\right)^2
\label{kincmass}
\end{equation}
where $\vec{V_{CM}}$ is the velocity of the centre of mass.
We run 5 different numerical experiments
where we keep the relative speed between the clumps ($140~km/s$) constant,
and we vary the speed of each clump, and consider maximum speeds of 
$70$ $km/s$, $80$ $km/s$, $90$ $km/s$, $100$ $km/s$, and $110$ $km/s$.
Fig.\ref{mapsasymv}a illustrates some of these simulations.
Fig.\ref{mapsasymv}b illustrates the evolution of the
kinetic energy of two such numerical experiments:
one in which both clumps have $70$ $km/s$ (blue - which corresponds to the numerical experiment
presented in Sec.~\ref{simcollision}), and one where the clumps have different speeds and are 
travelling at $110$ $km/s$ and $30$ $km/s$.
The evolution of the magnetic energy associated with the $y-$component
of the magnetic field, and therefore with the distortion of the field,
is fundamentally the same between both cases (within $3\%$).
We thus find that the amplitude of the kink mode induced by the collision
remains the same despite the changing speed between the two clumps.
We find differences of about $0.04$ $Mm$ for the peak value and
the evolution of the kink amplitudes largely overlap,
as the kinetic energy with respect to the centre of mass
is essentially the same in both simulations.

\subsection{Repartition of kink and sausage modes amplitudes}

\begin{figure}
\centering
\includegraphics[scale=0.30]{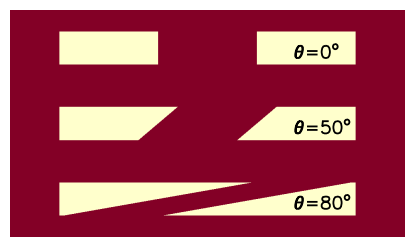}

\caption{Sketch to illustrate the initial conditions of the simulations
of the parameter space investigation where we vary the angle $\theta$
between the internal faces of the clumps and the 
direction perpendicular to the direction of travel.}
\label{mapsang}
\end{figure}

In this section, we address the geometry of the colliding clumps in order 
to study how this affects the excitation of kink and sausage modes along the waveguide.
We therefore vary the shape of the colliding clumps by 
changing the angle $\theta$ between the internal faces of the clumps and the 
direction perpendicular to the direction of travel. 
In particular, we consider
values of $\theta=[0^{\circ},20^{\circ},40^{\circ},
45^{\circ},50^{\circ},60^{\circ},80^{\circ}]$.
Fig.~\ref{mapsang} illustrates some of the configurations.
The numerical experiment with $\theta=50^{\circ}$ is the one presented in Sec.~\ref{simcollision}.
All the experiments in this section start with the same kinetic energy as the different clump shapes considered have the same area.

\begin{figure}
\centering
\includegraphics[scale=0.40]{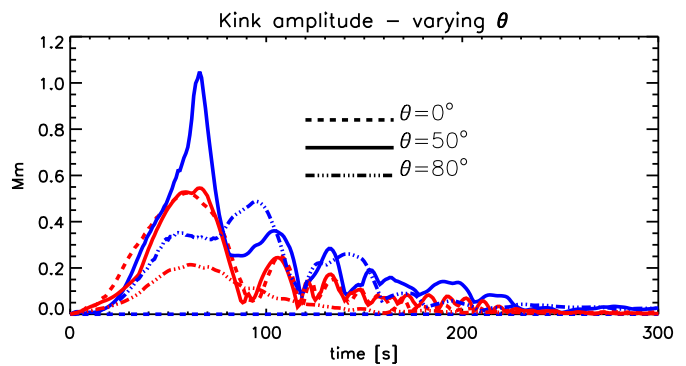}
\caption{Evolution of the functions $K(t)$ (blue lines) and $S(t)$ (red lines)
for some of the simulations where we vary the angle $\theta$
between the internal faces of the clumps and the 
direction perpendicular to the direction of travel.}
\label{modestime_ang}
\end{figure}

\begin{figure}
\centering
\includegraphics[scale=0.40]{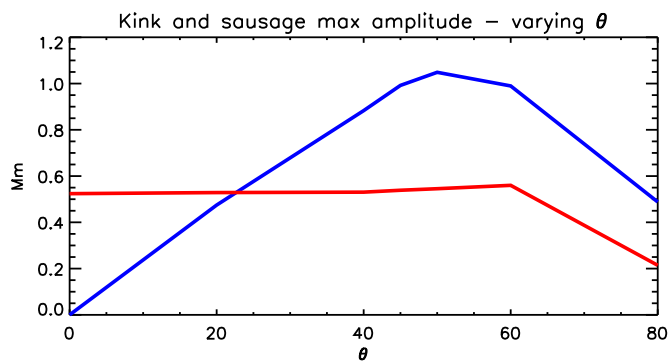}

\caption{Maximum of $K(t)$ (blue line) and $S_e(t)$ (red line) as a function of the angle $\theta$.}
\label{angmodes}
\end{figure}

For all simulations, we analyse the amplitudes of the excited kink and sausage modes
using the techinique outlined in Sec.\ref{ksmodes}.
Fig.~\ref{modestime_ang} shows the evolution of these amplitudes as a function of time
for 3 specific simulations
that illustrate 3 different regimes:
when the clumps are symmetric ($\theta=0^{\circ}$),
when the clumps are not symmetric ($\theta=50^{\circ}$),
and when the clumps are
significantly inclined ($\theta=80^{\circ}$).
Fig.~\ref{angmodes} shows the profiles
of the maximum kink and sausage amplitudes found in each numerical experiment
as a function of the angle $\theta$.

We find that when $\theta=0^{\circ}$,
only the sausage mode is excited and 
no kink is observed.
This result is a consequence of the perfect symmetry of the system
leading to no force imbalance in the $y$ direction and thus only symmetric oscillations can be triggered.
In the considered parameter space, the maximum of
the sausage mode amplitude is  about $0.5$ $Mm$, which is obtained for all considered values of $\theta\leq60^{\circ}$.
However, for $\theta\neq0$ a kink oscillation
is triggered as well. We also find that the 
oscillations after the collision follow a similar time evolution
for all visible kink and sausage modes for $\theta\leq60$ (Fig.~\ref{modestime_ang}).
This means that the period of the oscillations does not crucially depend on the shape
of the clumps, but on the local properties of the plasma,
i.e. the local Alfv\'en speed.

As seen in Fig.~\ref{angmodes}, the amplitude of the sausage mode does strongly depend on the shape of the clumps,
unless the collision angle is very large ($\theta>60$). When $\theta$ becomes
too large, the geometric configuration leads to smaller gas pressure gradients and 
consequently smaller magnetic field distortion
and both sausage and kink modes are not significantly excited.
This result is in accordance with our previous finding that the kinetic energy with respect to the centre of mass is a key parameter. Since the asymmetry introduced by changing the collision angle does not really affect the centre of mass, it is expected that the sausage mode will not be affected (i.e. similar standing sausage modes are generated). In contrast, the kink mode is sensitive to the internal structure of the collision
because it can lead to an imbalance of the forces along the $y$ direction.
Its amplitude steadily increases from $\theta=0^{\circ}$ to $\theta=50^{\circ}$,
as in this range the asymmetry of the collision increases
and the mechanism that triggers the kink mode becomes more efficient.
After $\theta=50^{\circ}$, the collision
leads to smaller gas pressure gradients,
as the clumps experience a weaker collision,
and less energy is transferred to kink modes.
As a result, an angle of $\theta\sim50^{\circ}$ seems to be the optimal 
inclination to trigger kink modes through counter-streaming flow collisions. 
At this inclination, the ratio between the kink and sausage amplitude is maximum
and it then remains roughly constant as increasing the clump inclination
further leads to weaker collisions
and both kink and sausage modes lose power.

\begin{figure}
\centering
\includegraphics[scale=0.30]{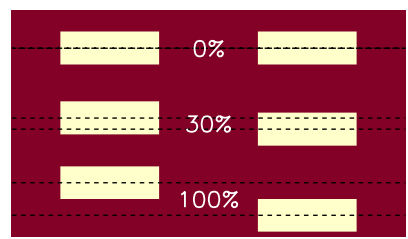}

\caption{Sketch to illustrate the initial conditions of the simulations
of the parameter space investigation where we vary the
offset between the colliding clumps.}
\label{mapsoff}
\end{figure}

Similarly, we investigate whether an initial offset in the travelling
direction of the clumps has the same asymmetric properties that
lead to the generation of both kink and sausage modes.
Fig.~\ref{mapsoff} illustrates this investigation,
where we consider two clumps with a flat front ($\theta=0^{\circ}$), 
but with 9 different values of offset between the centres of the clumps,
from $0\%$ (the two clumps on the same y coordinate), 
to $100\%$ (the two clumps displaced enough not to collide). 

\begin{figure}
\centering
\includegraphics[scale=0.40]{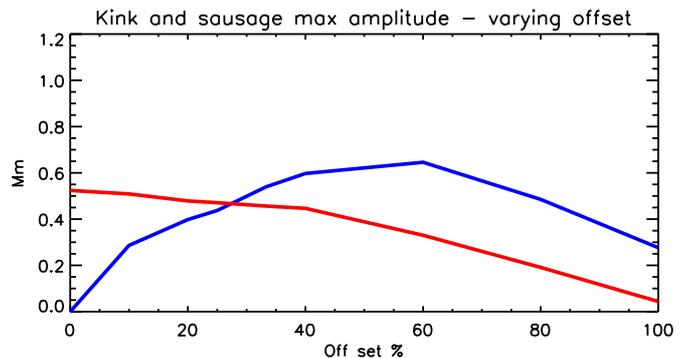}

\caption{Maximum of $K(t)$ (blue line) and $S(t)$ (red line) as a function of the offset between the colliding clumps.}
\label{offmodes}
\end{figure}

Fig.~\ref{offmodes} shows the maximum kink and sausage amplitudes found
in each simulation.
By comparing with Fig.~\ref{angmodes}, we find that this geometric configuration leads to 
generally smaller kink and sausage amplitudes.
However, some patterns are in common between varying the offset between the clumps
and varying the angle $\theta$ of the fronts. 
In Fig.~\ref{offmodes}, we find that the sausage amplitude steadily decreases
as the offset increases.
Again, this can be explained by the variation of the centre of mass, which results in a reduction of the colliding surface between the two clumps.
The behaviour of the kink mode is similar as before in that the offset produces a change in the forces along the $y$ direction. These forces reach a maximum for an offset of $\theta=60^{\circ}$ and decrease at higher offset values when the collision becomes weaker because of the limited interaction between the clumps.
As for the previous case, the ratio between kink and sausage modes initially increases 
with the asymmetry in the system and it then remains constant.

\subsection{Lengths and widths of the clumps}

\begin{figure}
\centering
\includegraphics[scale=0.40]{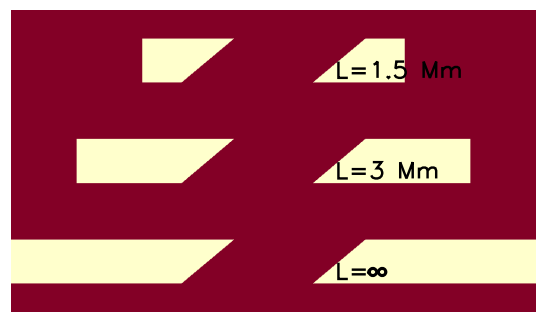}
\caption{Sketch to illustrate the initial conditions of the simulations
of the parameter space investigation where we vary the
length of the clumps.}
\label{mapslen}
\end{figure}

In this section, we investigate the role of the lengths and widths of the clumps in generating kink and sausage modes.
Fig.~\ref{mapslen} illustrates some of the experiments we run with a varying length
of the clumps.
We consider clumps with lengths
$L$ of 1.5~Mm, 2.4~Mm, 3~Mm, 3.6~Mm, 3.9~Mm,  4.2~Mm
and additionally another simulation where the initial clumps
are long enough to touch the $x$-boundaries of the simulation box.
This arrangement makes this simulation equivalent to having infinitely long clumps
because of the boundary conditions. This scenario may effectively correspond to very long clumps, as sometimes observed in coronal rain \citep{Antolin2015}.

\begin{figure}
\centering
\includegraphics[scale=0.40]{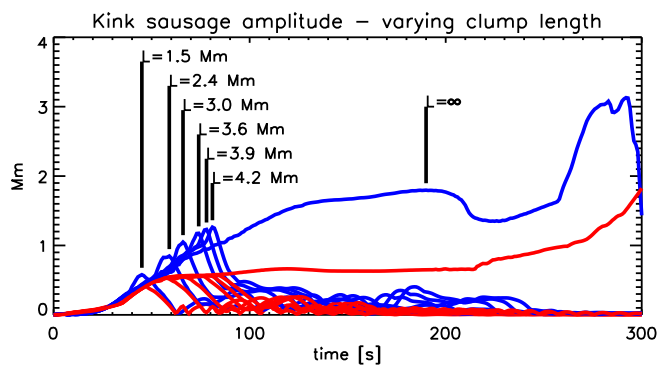}

\caption{Evolution of the functions $K(t)$ (blue lines) and $S(t)$ (red lines)
for some of the simulations where we vary the length of the clumps.} 
\label{lenmodes}
\end{figure}

Fig.~\ref{lenmodes} shows the measured amplitudes for the kink and sausage modes as
a function of time for these simulations.
The basic evolution is common to both kink and sausage modes. 
The amplitude increases as long as the clumps are colliding 
and this period lasts longer proportionally to the clump's length. Thus the maximum is reached at different times.
At the same time, the amplitude increases as the collision leads to an increased distortion of the magnetic field.
We find that this distortion, i.e. the maximum kink amplitude, initially scales linearly with the kinetic energy, but we observe some saturation effects from $L=3.6~Mm$.
The saturation occurs because the magnetic tension force
is proportional to the distortion of the magnetic field 
and at some point the resulting restoring force becomes enough to stop the magnetic field distortion.
This is also seen in the simulation with infinite lengths.
The kink amplitude keeps increasing and it then saturates
shortly after $t=100$ $s$. However, for this case we observe a permanent distortion of the field, and an additional stronger distortion at the end.
\begin{figure}
\centering
\includegraphics[scale=0.40]{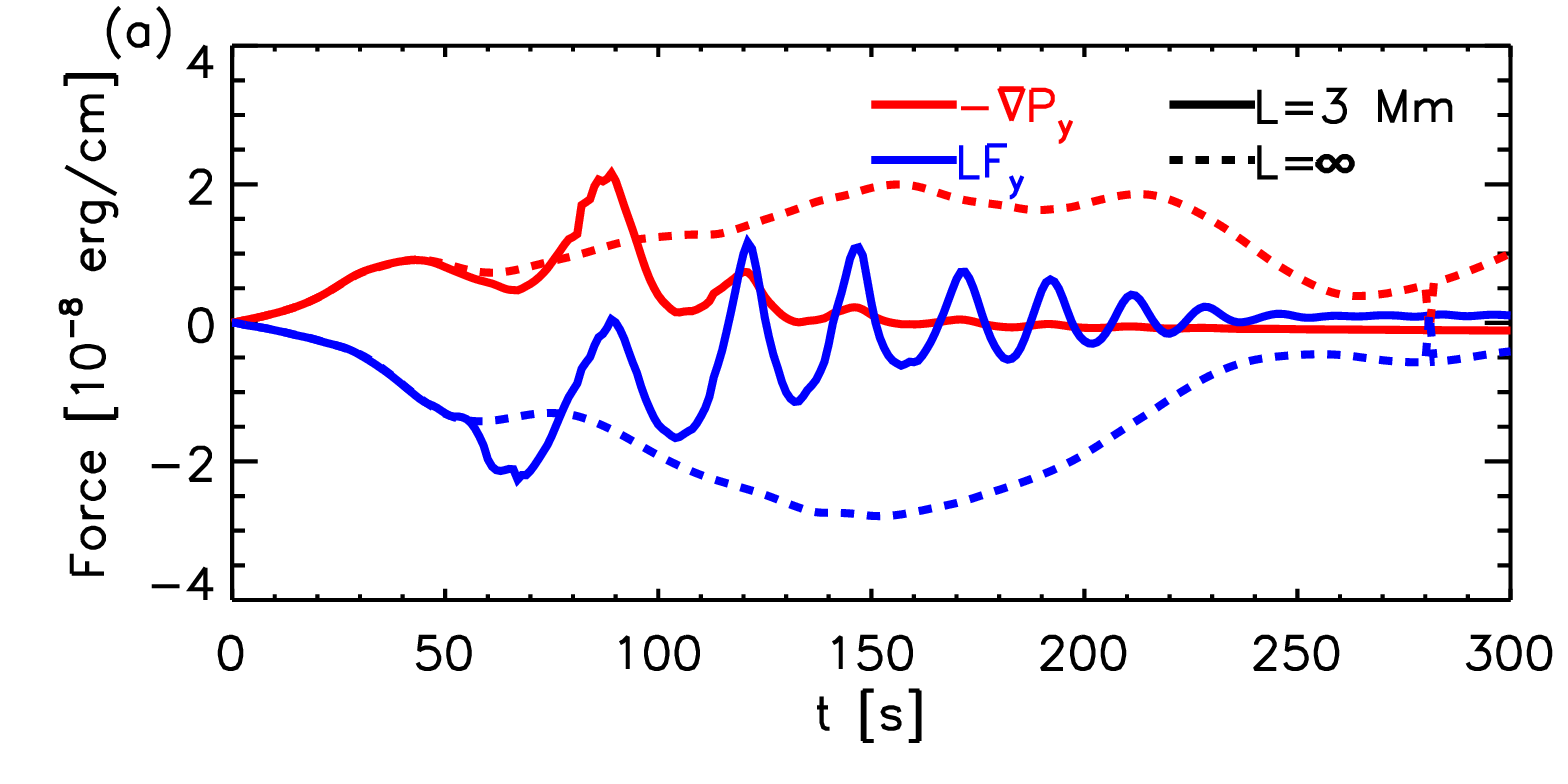}

\includegraphics[scale=0.40]{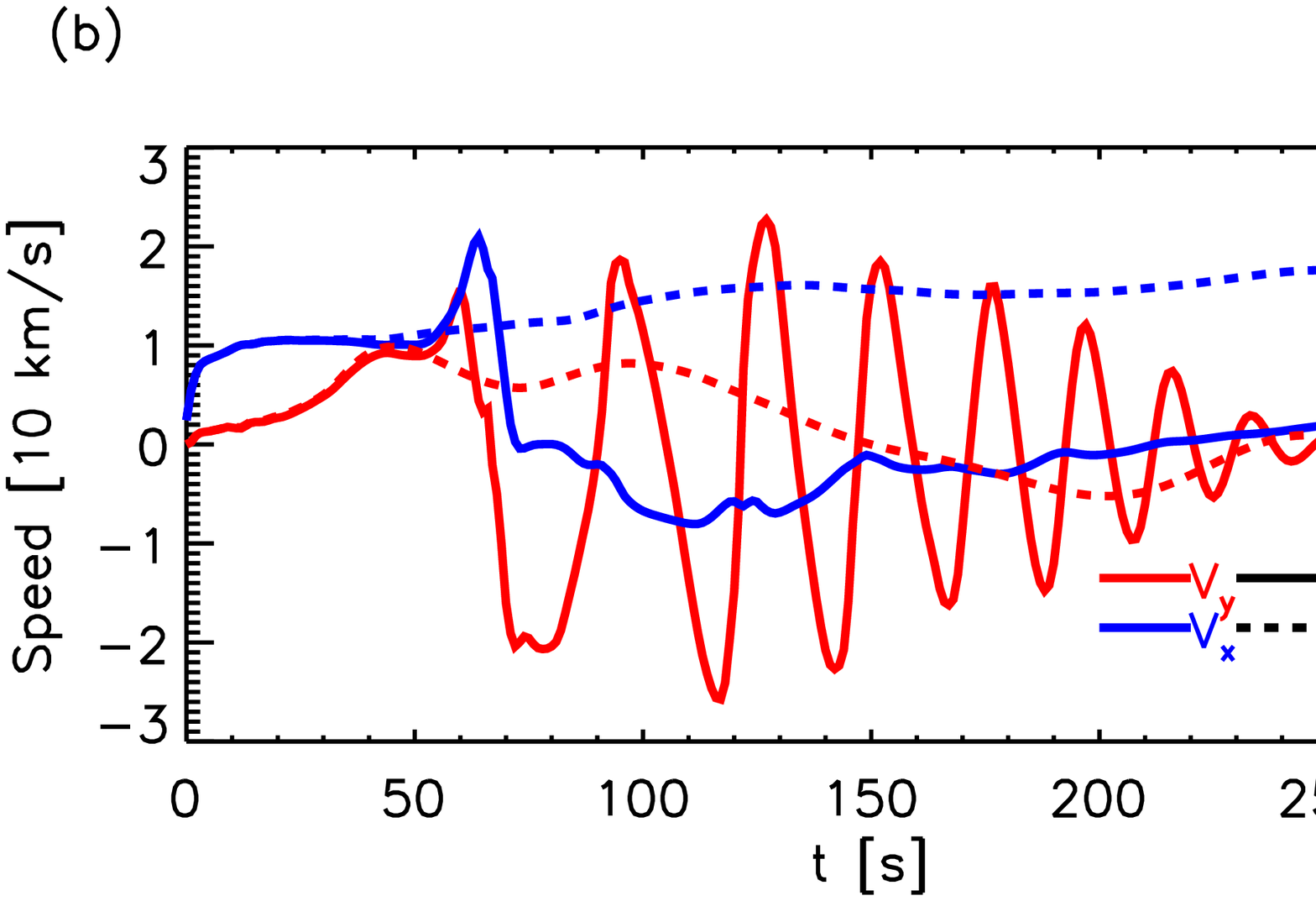}

\includegraphics[scale=0.30]{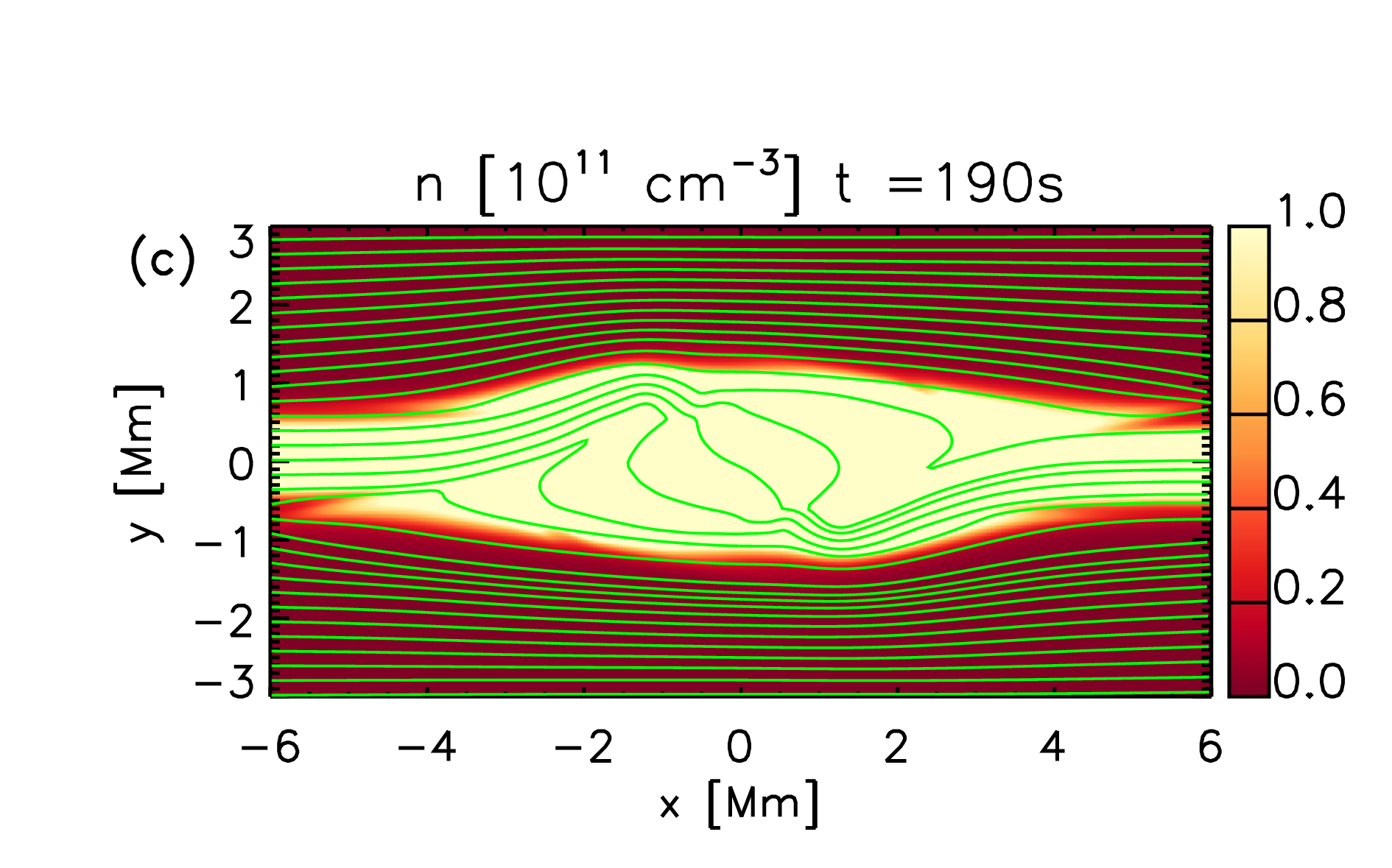}
\caption{(a) Evolution of the $y$-components of the gas pressure gradient and Lorentz force averaged over the rectangular region shown in Fig.~\ref{snaps}
for the simulations with clump length $L=3$ $Mm$ and infinite.
(b) Evolution of the $x$ and $y$ components of the velocity averaged over the rectangular region shown in Fig.~\ref{snaps}
for the simulations with clump length $L=3$ $Mm$ and infinite.
(c) Map of the number density $n$ at $t=190$ $s$ for 
the simulation with infinite clump length.
Green lines are magnetic field lines drawn from the left-hand side boundary.}
\label{lenforces}
\end{figure}

Fig.~\ref{lenforces}a compares the forces (gas pressure gradient and Lorentz force)
that develop in the simulation where $L=3$ $Mm$ and $L=\infty$
averaged over the rectangular region defined in Fig.~\ref{snaps}.
The evolution of the forces for all the simulations with a finite clump length
is similar and they only differ in the time when the magnetic restoring forces 
start driving the system.
The evolution of the forces is qualitatively different
when we consider infinite clumps as the gas pressure gradient does not undergo the
stages we described in Sect.~\ref{simcollision} and the restoring magnetic force 
grows until $t=150$ $s$ without showing an oscillatory behaviour afterwards.
Similarly, Fig.~\ref{lenforces}b shows the average $v_x$ and $v_y$ over the same region.
This evolution also confirms that when the clumps are infinite no oscillatory behaviour sets in on these spatial and temporal scales. 
Moreover, the average $v_x$ remains always positive, effectively leading to
counter-propagating streams, one positive flowing in the region $y>0$
and one negative flowing in the region $y<0$.

\begin{figure}
\centering

\includegraphics[scale=0.40]{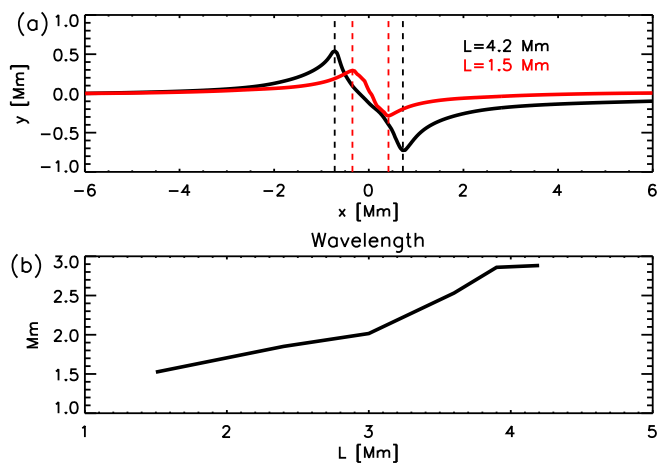}
\caption{(a) Profile of $D(x)$ at the time of maximum of $K(t)$
for the simulations with $L=1.5$ $Mm$ and $L=4.2$ $Mm$.
(b) Characteristic wavelength of $D(x)$ at the time of maximum $K(t)$
as a function of the length of the clumps.}
\label{lenmodes_lambda}
\end{figure}

Fig.~\ref{lenforces}c shows the system at $t=190$ $s$
for the simulation with infinite clumps,
when the saturation is already well established.
During the saturation phase, the distortion of the magnetic field remains constant
in terms of the magnetic tension force, but it travels
away from $x=0$. When it reaches the boundaries
the kink mode propagation is affected by the numerical boundary conditions
and it alters the obtained stationary state.
This can be understood by the fact that the constant addition of mass through the boundaries constitutes a constant addition of kinetic energy for the collision, which then forces the system to slowly reach a new state of equilibrium.
At the same time, it should be noted that such a configuration when evolving in the presence of non-ideal MHD effects (such as magnetic field diffusion), could lead to the formation of an isolated plasmoid if the density structure gets fragmented
where the magnetic field reconnects.
Fig.\ref{lenforces}c shows a few locations where the magnetic field changes orientation and the magnetic reconnection could lead to the detachment of portions of dense clumps. This additional perturbation 
could lead to the generation of further MHD waves.
Additionally, more energetic collision from shorter density clumps can lead to the same magnetic field distortion,
as we found in the numerical experiments with highest kinetic energy in Table~\ref{tablerhocvc}.
Another numerical experiment that presents some similarities with our approach has been performed by \citet{2016ApJ...833...36F}, in which the combination of collision and shear flow leads to the generation of Kelvin-Helmholtz instabilities and the onset of magnetic reconnection. Future studies will address the role of non-ideal MHD terms in this phenomenon.

Fig.~\ref{lenmodes_lambda}a shows the kink profile for two simulations
with $L=1.5$ $Mm$ and $L=4.2$ $Mm$ at the respective times when the kink oscillations
are maximal.
We find that the amplitude of the kink is larger for $L=4.2$ $Mm$
and it also presents 
a larger distance between the two peaks of the oscillation.
This happens because as soon as the collision starts, these peaks
travel away from $x=0$ at the local Alfv\'en speed.
The distance between the two peaks can be thought of as half the wavelength of the kink mode.
In Fig.\ref{lenmodes_lambda} we plot this wavelength
as a function of the clump's length (excluding $L=\infty$).
We find that the wavelength linearly depends on the clump's length
for the range we investigate in this study.
This can be understood by the analysis of forces conducted in Sec.~\ref{simcollision}.

\begin{figure}
\centering
\includegraphics[scale=0.40]{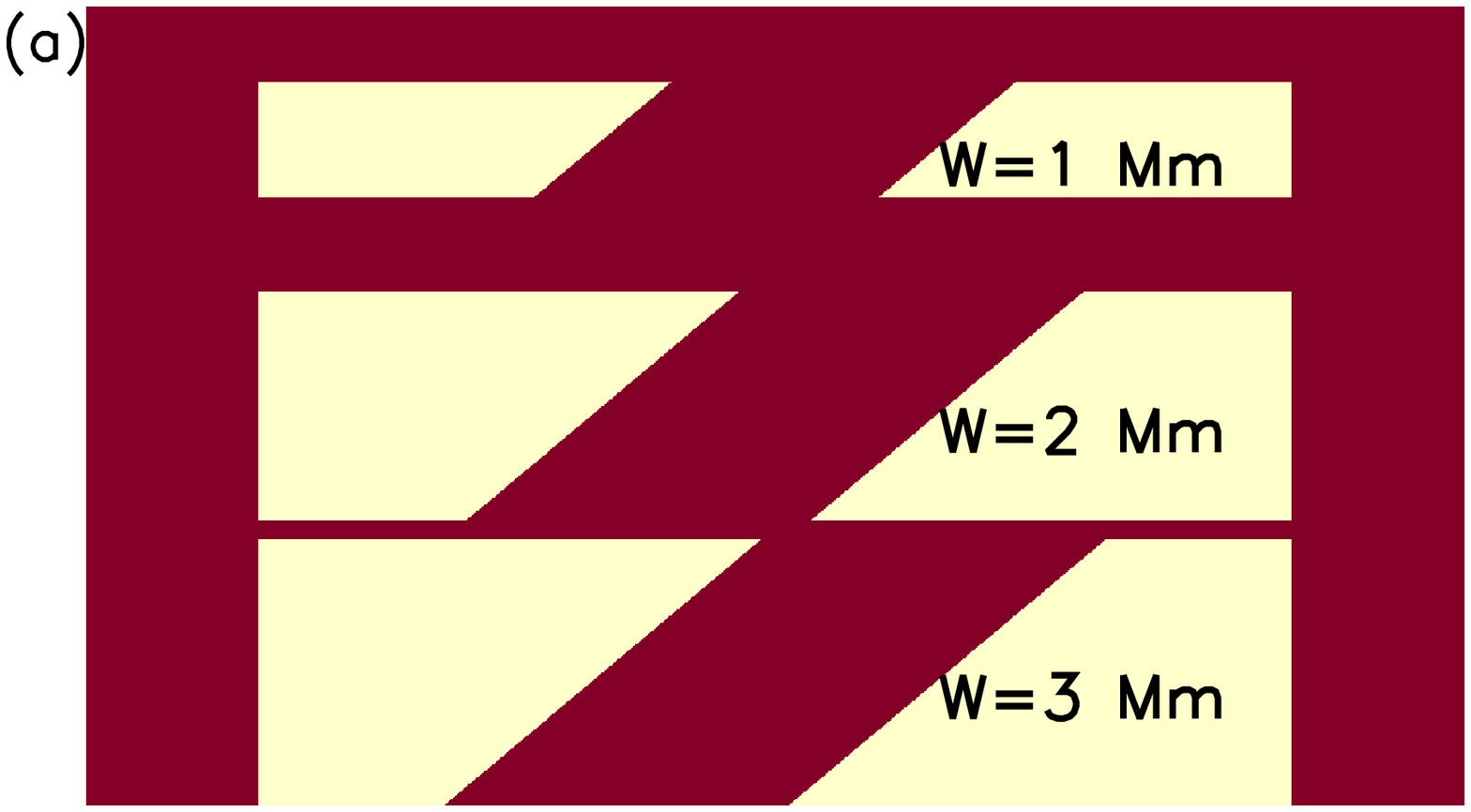}
\includegraphics[scale=0.40]{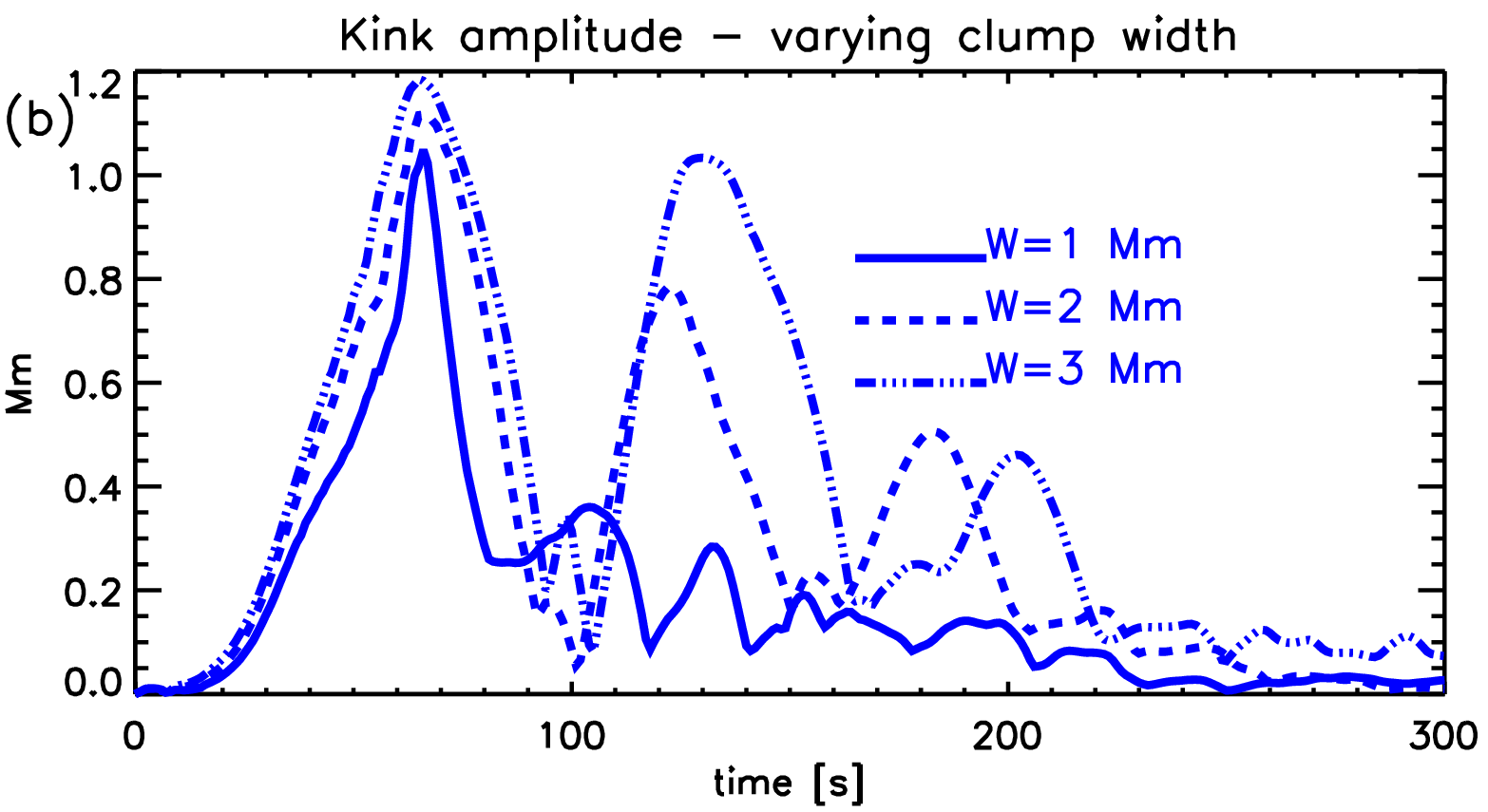}
\caption{(a) Sketch to illustrate the initial conditions of the simulations
of the parameter space investigation where we vary the
width of the clumps.
(b) Evolution of the functions $K(t)$ 
for the simulations where we vary the width of the clumps.}
\label{widmodes}
\end{figure}

Finally, we study the effect of increasing the width of the clumps.
Fig.~\ref{widmodes} shows different clump configurations
when we vary the clumps' widths (keeping both widths equal).
Since their lengths do not change
we do not expect a significant change in the wavelength of the 
generated kink oscillations.
We investigate three different cases with (symmetric) width values $W$ of 1~Mm, 2~Mm, and 3~Mm.
Fig.~\ref{widmodes} shows the kink amplitude 
as a function of time for these three experiments.
We find that the maximum amplitude is still linearly dependent on the initial kinetic energy,
but as the width of the clumps increases, the post-collision oscillations 
show less damping,
as the wave guide becomes larger as well.
The damping, i.e. the ratio between the amplitude of the first two peaks of the oscillation in Fig.~\ref{widmodes}
ranges between $0.34$ for the simulation with $W=1$ $Mm$
to $0.87$ for the simulation with $W=3$ $Mm$.
In contrast, for the sausage modes amplitude (not shown here),
the damping does not significantly depend on the clumps width.

\section{3D simulations}
\label{3dsimulation}

Finally, we consider the collision of two clumps
and the subsequent excitation of MHD waves 
in a 3D configuration in order to understand
how this mechanism occurs in a 
more realistic geometrical configuration.

\begin{figure*}
\centering
\includegraphics[scale=0.28]{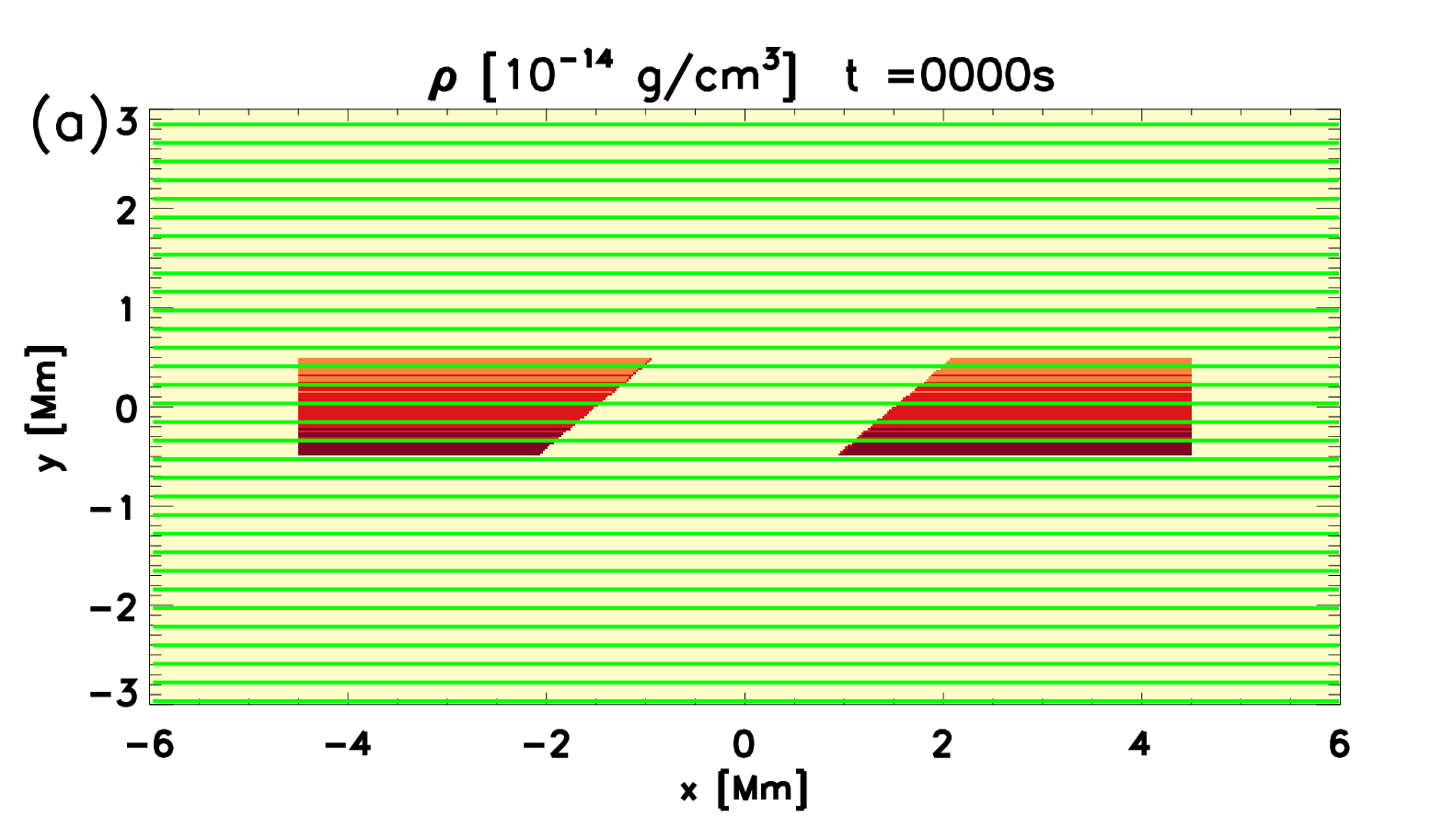}
\includegraphics[scale=0.28]{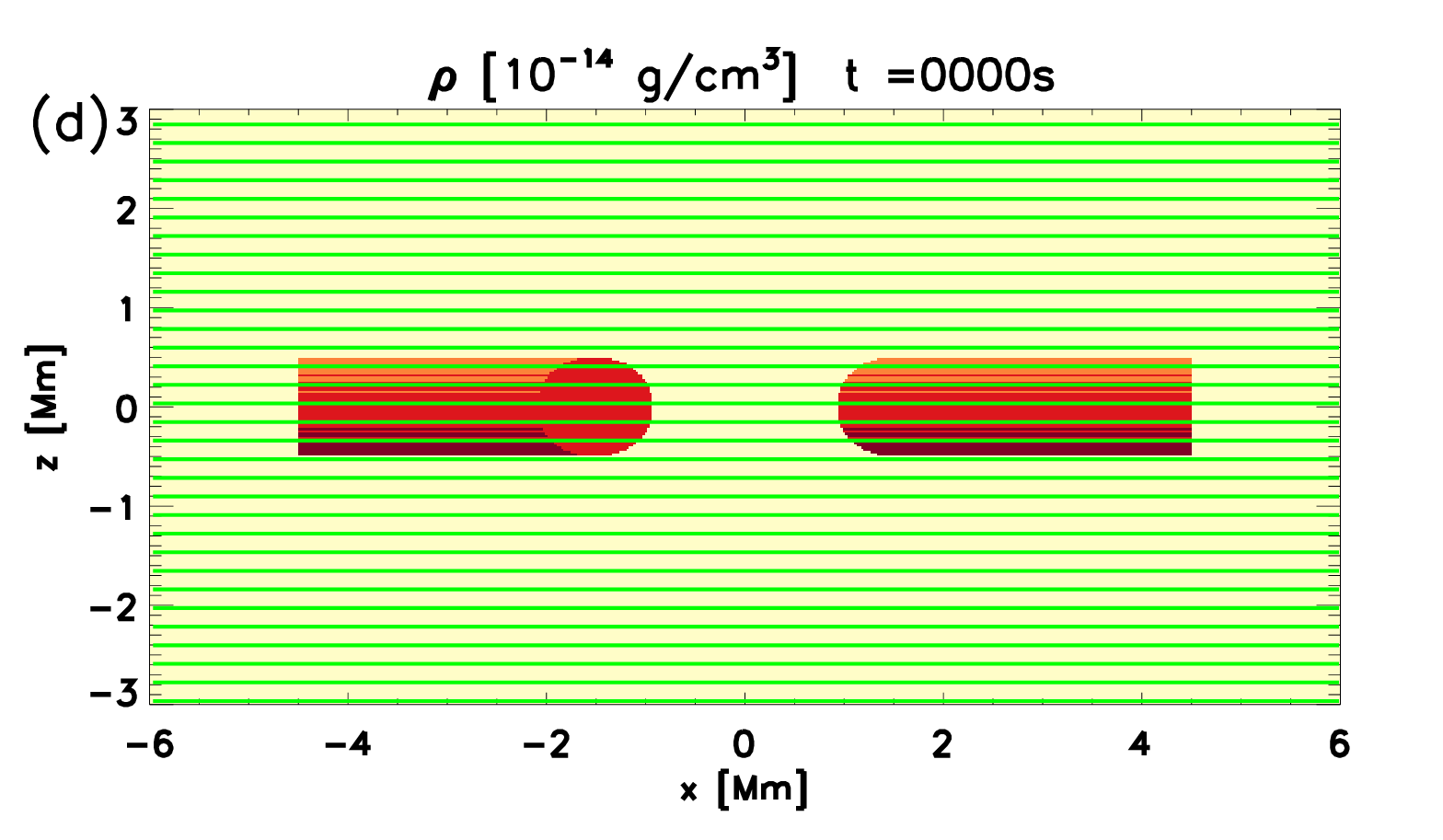}

\includegraphics[scale=0.28]{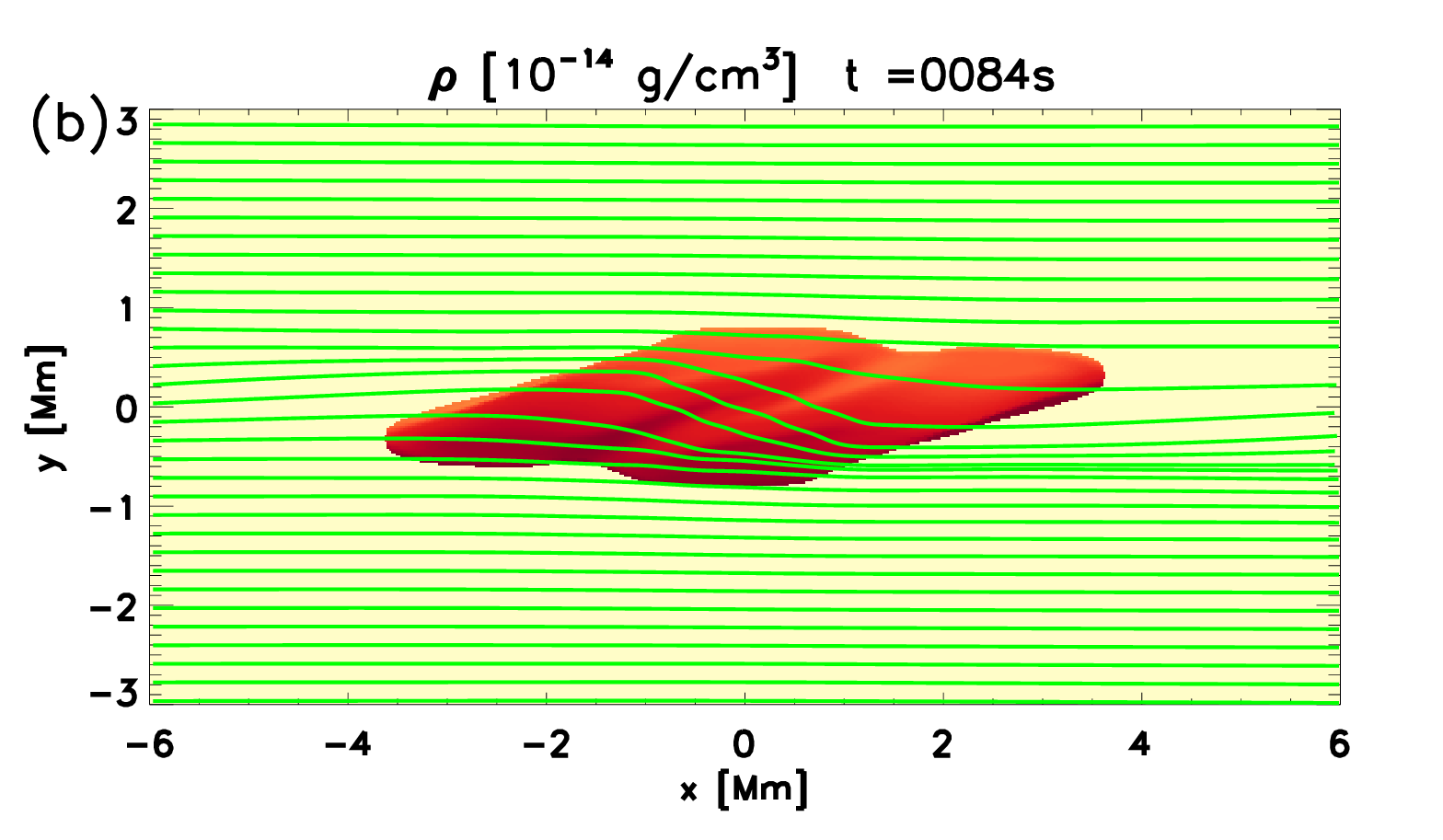}
\includegraphics[scale=0.28]{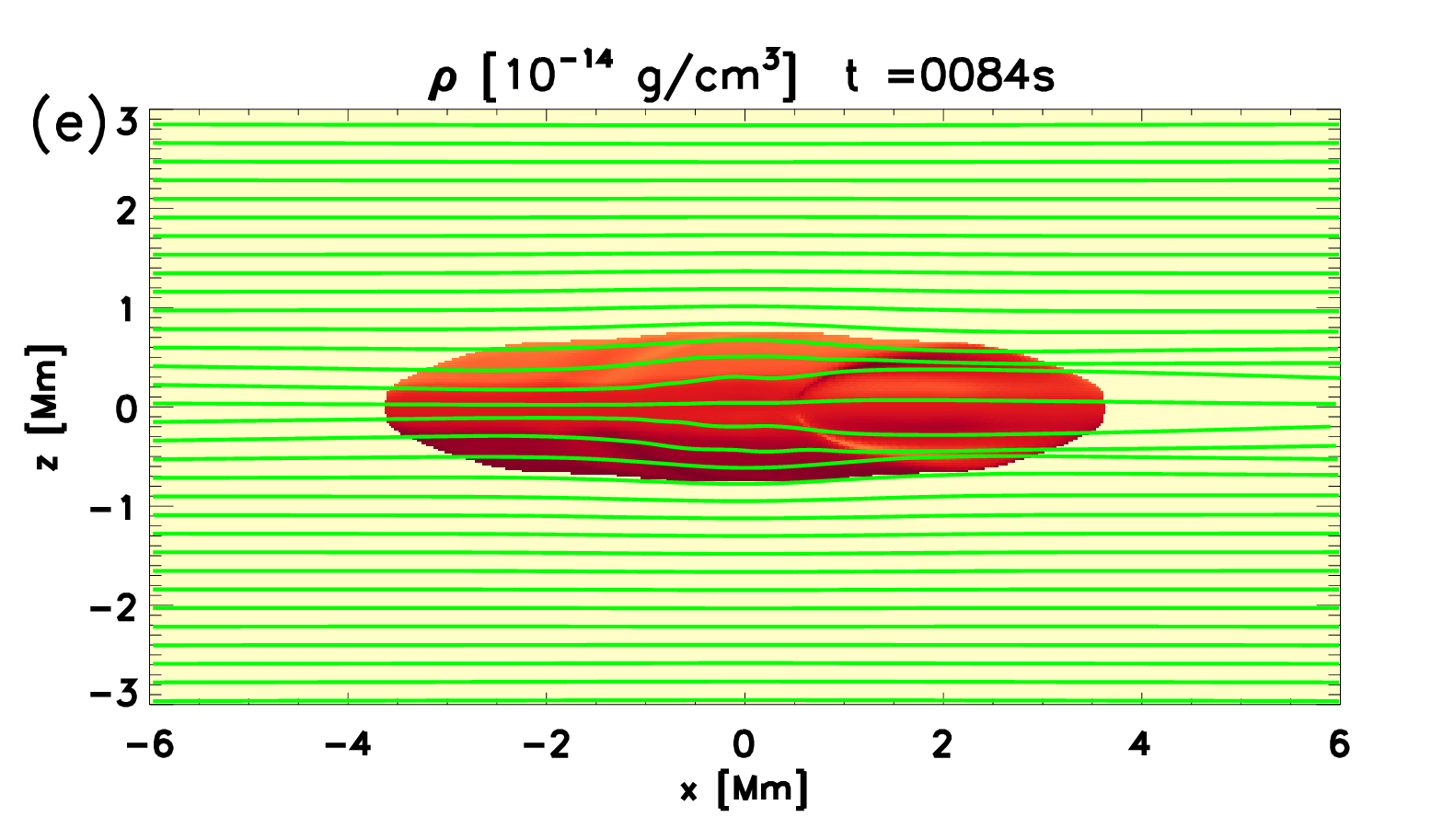}

\includegraphics[scale=0.30]{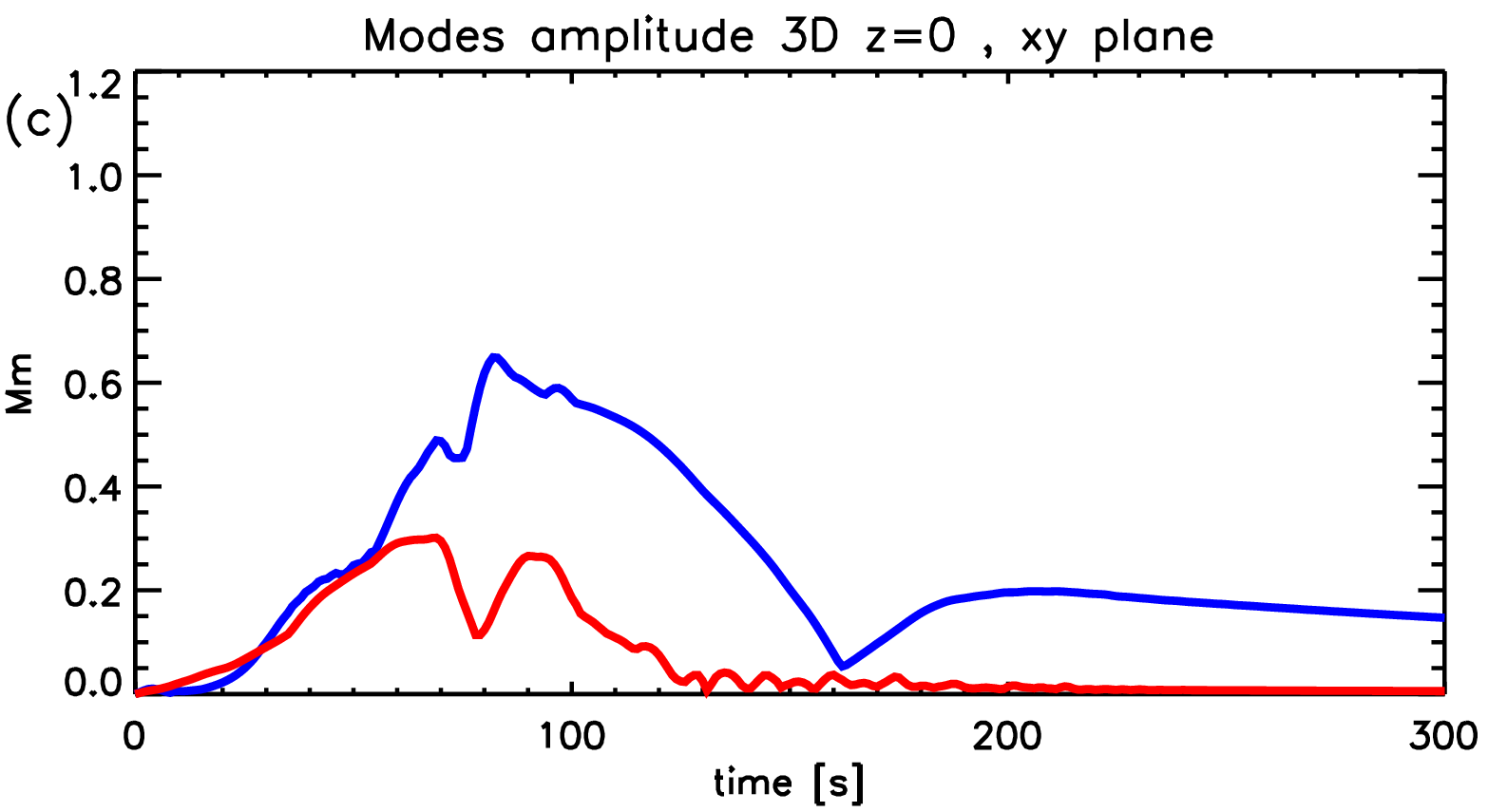}
\includegraphics[scale=0.30]{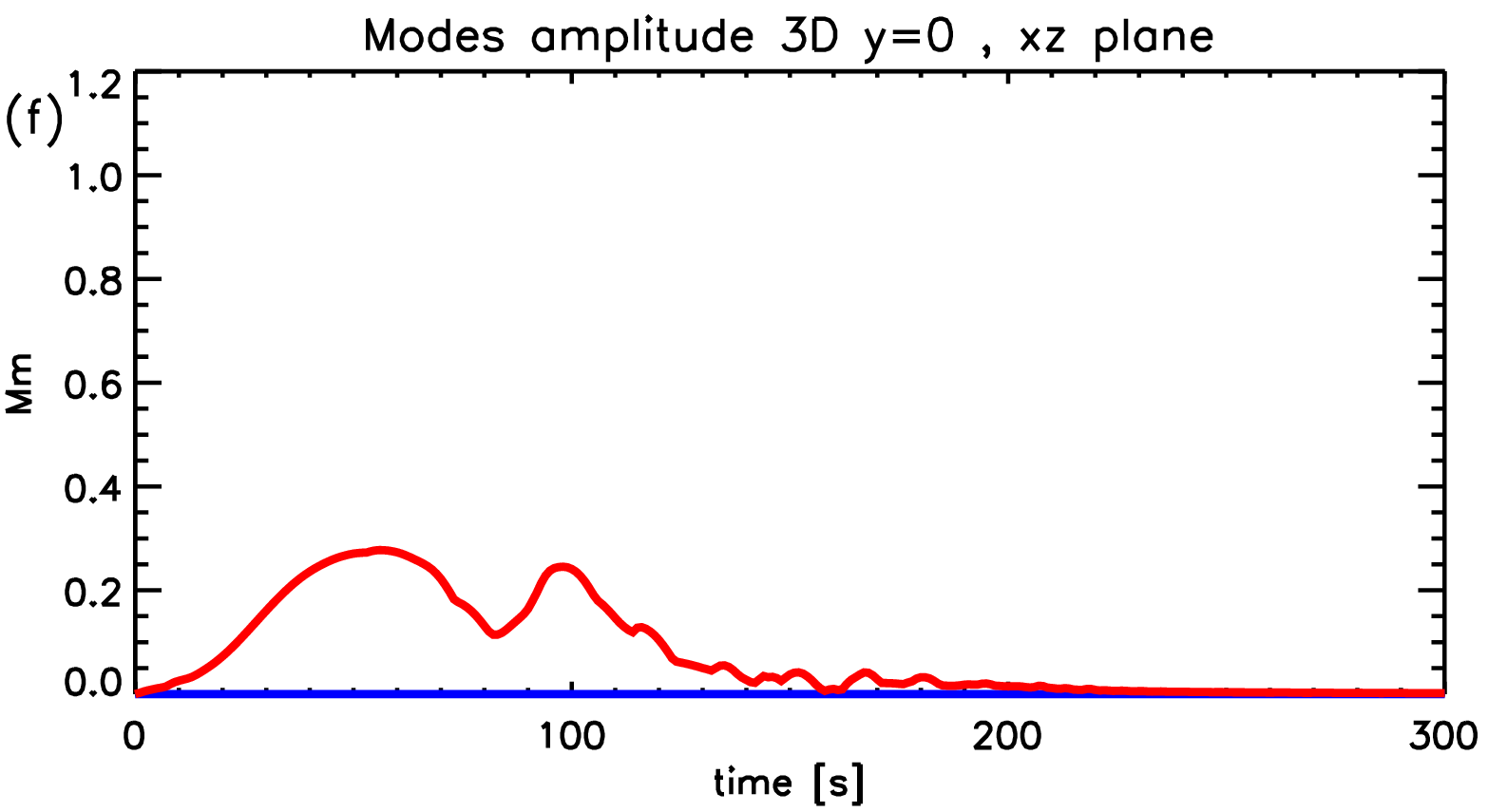}

\caption{
(a) and (d) Initial density contour projected on the xy-plane 
and on the xz-plane, respectively. (b) and (e) Same contours at $t=84$ $s$.
Green lines are magnetic field lines drawn from the left-hand-side boundary.
Evolution of the functions $K(t)$ (blue lines) and $S(t)$ (red lines)
for the 3D MHD simulation
on the z=0 (c) and y=0 planes (f).
Movie available online.}
\label{3drho}
\end{figure*}

The two clumps now
have a cylindrical shape where the facing
surfaces are still oblique.
Panels a and d of Fig.~\ref{3drho}
show the initial condition of this simulation from two perpendicular viewpoints.
The numerical setup is necessarily changed with respect
to the 2D simulations, as we take into consideration
the z direction extending from $-3~Mm$ to $3$ $Mm$.
The normal to the colliding faces of the clumps is parallel to the $xy-$plane, and therefor does not have a $z-$component. Moreover the spatial resolution of the 3D simulation is 
a factor of 2 coarser than the one in the 2D simulation.

The main difference when addressing the problem in a 3D geometry is
that the diagnostic of its evolution crucially depends on the viewpoint.
In the $z=0$ plane the configuration is analogous to the 2D experiments we have
already investigated, but when we consider the $y=0$ plane, the clumps collision appear symmetric.
Moreover, we expect that the thermal pressure gradient between the colliding clumps will exert a force in
all directions perpendicular to $\vec{B}$
and this will result in a smaller deformation
of the magnetic field in all directions.
Panels d and e of Fig.~\ref{3drho} show the evolution
of the system after $84$ $s$, a time at which
the clumps have merged after the collision.
At this time, the plasma of the dense clumps
has compressed the magnetic field after moving past the collision region.
At a later stage, the clumps' plasma fill the whole wave guide.

In order to analyse the kink and sausage modes
induced in this case, we use the same technique introduced in Sect.~\ref{ksmodes}. However, as this is an inherently 2D technique we apply it
to the $z=0$ plane and to the $y=0$ plane separately.
These two different cross-sections cut the system perpendicularly 
along the directions where the asymmetry is most pronounced
and where the system becomes symmetric.
Fig.~\ref{3drho} shows the kink and sausage modes evolution for both planes.

The evolution along the $z=0$ plane  is more similar to the 2D scenario, showing the development of both kink and sausage modes. 
As expected, the amplitude of the kink modes is smaller than in the 2D configuration
as the deformation occurs in a larger volume.
The maximum kink amplitude found here is about 60\% of the one found in the corresponding 2D simulation.
Additionally, the oscillatory behaviour of the kink mode amplitude is
less evident in 3D. The maximum of the kink mode amplitude is reached at $t=82$ $s$,
and afterwards it steadily decreases until $t=160$ $s$ when it
reaches a minimum before growing again,
because of an apparent inclination of the waveguide,
due to the magnetic field lines being tracked starting from one boundary.
The kink modes are almost completely damped 
within the first oscillation and
this evolution is fundamentally different than the one observed in 2D, where
we identify at lease three kink mode oscillations peaks in the kink modes amplitude.
The kink modes are not visible when we move to the $y=0$ plane
as the system becomes symmetric there.
The amplitudes of the sausage modes are instead very similar for both
$z=0$ and $y=0$ planes as the nature of this deformation does not depend 
on the asymmetry (as already pointed out previously) and these are also about 60\%
of the amplitude found in the corresponding 2D simulation.
In both planes, the sausage mode amplitude initially increases because of the collision as we have already shown, but it then shows a second peak
at $t\sim100$ $s$. This second peak is due to the compression of the 
magnetic field lines when the clumps travel past the collision region.

\section{Discussion and Conclusions}
\label{conclusions}

In this paper, we have investigated the collision of counter propagating clumps
(density enhancements), a mechanism
that can generate transverse MHD waves in-situ in the solar corona.
Such mechanism has been observed in action in coordinated observations between \textit{SDO}, \textit{IRIS}, and \textit{Hinode}
of a prominence/coronal rain complex, as presented in paper I.

The main focus of this study is to use
MHD simulations to analyse in detail the generation
of kink and sausage modes by means of such collisions. 
Hence, we have analysed in detail a
2D ideal MHD simulation that could explain the observations
and we have then performed an extended parameter space investigation.
In this paramater space investigation we have varied
both the clumps properties and the overall geometry,
running also a 3D MHD ideal simulation of the same setup.

In order to analyse our results, we developed a simple technique to identify
the amplitude of kink and sausage modes in a wave guide. 
We associate the kink amplitude with the asymmetric distortion
of magnetic field lines that are internal to the clumps
and the sausage amplitude to the symmetric displacement of
the magnetic field lines enveloping the collision region.

We find that the collision of two counter propagating clumps 
is a viable way to generate kink and sausage modes.
In particular, we have identified two main phases in this process.
The first one is the collision of the clumps during which
the dynamics are dominated by the motion and merging of the dense plasma.
In this phase, trains of fast MHD waves travel outwards from the collision region.
During this time, the magnetic field is distorted in a way that crucially depends
on the clump's length and shape. Any asymmetry between
the clumps will lead to the formation of kink modes,
while the formation of sausage modes is largely unaffected by the symmetry of the system. This phase terminates when the clumps have lost their kinetic energy, after which the restoring magnetic forces (mostly magnetic tension)
that are proportional to the displacement of the magnetic field lines from their equilibrium
configuration, trigger the oscillations that propagate along the wave guide.

By varying the physical properties of the colliding clumps, 
we established that the amplitude of the kink and sausage modes
is mostly dependent on the kinetic energy of the system with respect to the centre of mass.
At the same time, kink and sausage mode oscillations are more efficiently excited in different geometric configurations.
By varying the angle of the colliding fronts
or the overlap of the fronts,
we found that the amplitude of the sausage modes does not depend on the symmetry of the system,
but the interaction
between clumps is significantly reduced in extreme cases leading to less energy
in the resulting kink and sausage modes.
Intermediate regimes have the kink amplitude
always larger than the sausage mode amplitude. 
Moreover, the amplitude of the kink modes increases linearly with the clump's length because of the associated kinetic energy. However, very long clumps can reach a saturation regime.
This occurs when the restoring magnetic forces
due to the deformation of the magnetic field become equal to the 
thermal pressure gradient driven by the collision. 
The initial wavelength of the kink modes 
increases with the length of the clumps. The generated kink modes propagate away from the collision, while the generated sausage modes become standing due to the symmetry in the longitudinal forces. 
Finally, we found that while the initial kink amplitude is proportional to the kinetic energy, less damping is obtained for larger clump widths in our model.

In order to better relate with a realistic scenario, we extended our analysis to a fully 3D configuration where the clumps are cylindrical.
The key differences
are  i) the kink amplitude is nearly halved because the magnetic field lines distortion occurs on the volume around the cylinders
and, ii) the observed mode amplitudes and the apparent asymmetry of the system crucially depend on the viewpoint. However, only through forward modelling can this effect be properly estimated.

In conclusion, this work has improved our understanding of the mechanism
behind the generation of MHD waves in the solar corona due to the collision 
of counter propagating plasma clumps.
At the same time, more work is required in order to link this model with observations and future efforts will focus on forward modelling more realistic numerical models for proper comparison with observations 

\begin{acknowledgements}
This research has received funding from the UK Science and Technology Facilities Council (Consolidated Grant ST/K000950/1) and the European Union Horizon 2020 research and innovation programme (grant agreement No. 647214). P.A. acknowledges funding from his STFC Ernest Rutherford Fellowship (No. ST/R004285/1). This research was supported by the Research Council of Norway through its Centres of Excellence scheme, project number 262622.
This work used the DiRAC@Durham facility managed by the Institute for Computational Cosmology on behalf of the STFC DiRAC HPC Facility (www.dirac.ac.uk). The equipment was funded by BEIS capital funding via STFC capital grants ST/P002293/1, ST/R002371/1 and ST/S002502/1, Durham University and STFC operations grant ST/R000832/1. DiRAC is part of the National e-Infrastructure.
We acknowledge the use of the open source (gitorious.org/amrvac) MPI-AMRVAC software, relying on coding efforts from C. Xia, O. Porth, R. Keppens.

\end{acknowledgements}

\bibliographystyle{aa}
\bibliography{ref}

\end{document}